\documentclass[acmtog]{acmart}
\pdfoutput=1 
\AtBeginDocument{%
  \providecommand\BibTeX{{%
    \normalfont B\kern-0.5em{\scshape i\kern-0.25em b}\kern-0.8em\TeX}}}

\setcopyright{acmcopyright}\acmJournal{TOG}
\acmYear{2020}\acmVolume{39}\acmNumber{6}\acmArticle{199}\acmMonth{12} \acmDOI{10.1145/3414685.3417849}

\usepackage{overpic}
\usepackage{breakcites}
\usepackage{microtype}
\usepackage[ruled,linesnumbered]{algorithm2e}
\usepackage{wrapfig}
\usepackage{subfig}
\usepackage{pgfplots}
\usepackage{xcolor}
\usepackage{amsmath}
\usepackage{pict2e}
\usepackage{comment}
\usepackage{cancel}
\usepackage{graphicx}
\usepackage{soul}
\usepackage{float}

\usepackage[tableposition=above]{caption}
\usepackage[normalem]{ulem}
\graphicspath{ {./images/} }

\renewcommand{\sout}[1]{\ignorespaces}

\renewcommand{\textcolor}[2]{#2}
\newcommand{\M}{\mathcal{M}}

\usepackage{color}
\newcommand{\MM}[1]{{\color{red} {$ ^{MM}\{$ #1 $\}$}}}

\pgfplotsset{compat=newest}
\pgfplotsset{plot coordinates/math parser=false}
\newlength\figureheight
\newlength\figurewidth 
\usetikzlibrary{arrows.meta}
\acmSubmissionID{354}
\newtheorem*{remark}{\textbf{Remark}}
\newtheorem{defn}{\textbf{Definition}}
\newtheorem{thm}{\textbf{Theorem}}
\newtheorem*{thm2}{\textbf{Theorem}}
\newcommand{\RN}[1]{%
  \textup{\uppercase\expandafter{\romannumeral#1}}%
}
\begin{document}

\title{Nonlinear Spectral Geometry Processing via the TV Transform}

\author{Marco Fumero}

\email{fumero@di.uniroma1.it}
\affiliation{%
  \institution{Sapienza University of Rome}%
  \streetaddress{}
  \city{}%
  \state{}%
  \postcode{}%
}%

\author{Michael M\"{o}ller}

\email{Michael.Moeller@uni-siegen.de}
\affiliation{%
  \institution{Siegen University}%
    \streetaddress{}%
  \city{}%
  \state{}%
  \postcode{}%
}%
\author{Emanuele Rodol\`{a}}

\email{rodola@di.uniroma1.it}
\affiliation{%
  \institution{Sapienza University of Rome}%
    \streetaddress{}%
  \city{}%
  \state{}%
  \postcode{}%
}%

\begin{abstract}
We introduce a novel computational framework for digital geometry processing, based upon the derivation of a nonlinear operator associated to the total variation functional. Such an operator admits a generalized notion of spectral decomposition, yielding a convenient multiscale representation akin to Laplacian-based methods, while at the same time avoiding undesirable over-smoothing effects typical of such techniques. Our approach entails accurate, {\em detail-preserving} decomposition and manipulation of 3D shape geometry while taking an especially intuitive form: non-local semantic details are well separated into different bands, which can then be filtered and re-synthesized with a straightforward linear step. Our computational framework is flexible, can be applied to a variety of signals, and is easily adapted to different geometry representations, including triangle meshes and point clouds. We showcase our method through multiple applications in graphics, ranging from surface and signal denoising to enhancement, detail transfer, and cubic stylization. 
%
\end{abstract}




\maketitle
\section{Introduction}
In the last decades, the computer graphics community has witnessed a thriving trend of successful research in computational spectral geometry. Spectral approaches have met with considerable traction due to their generality, compact representation, invariance to data transformations, and their natural interpretation relating to classical Fourier analysis. Despite these benefits, the vast majority of current spectral methods rely upon the construction of smooth basis functions (obtained, for instance, as the eigenfunctions of the Laplace operator), leading to significant detail loss whenever the signal to be represented is not smooth. When this signal represents geometric information (e.g., the $(x,y,z)$ coordinates of mesh vertices), this inevitably leads to a ``lossy'' geometric encoding, \sout{which is} typically manifest as over-smoothing, vertex collapse, or spurious oscillations (so called Gibbs phenomenon) not appearing in the original geometry.

Such artifacts derive from the \sout{very} motivations underlying the seminal work of Taubin~\shortcite{Taubin:1995} and follow-up \cite{Desbrun:1999}, which were posed as an extension of the diffusion techniques from image processing to mesh {\em smoothing} applications. Today, the key driver behind their continued adoption lies in the convenient representation, which enabled significant gaps in performance in many relevant problems (e.g., shape correspondence \cite{ovsjanikov2012functional,zoomout}, vector field processing \cite{brandt2017spectral} among many others), while still suffering from often undesirable effects.

\begin{figure}[t!]
\begin{center}
  \begin{overpic}
  [trim=0cm 0cm 0cm 0cm,clip,width=1.0\linewidth]{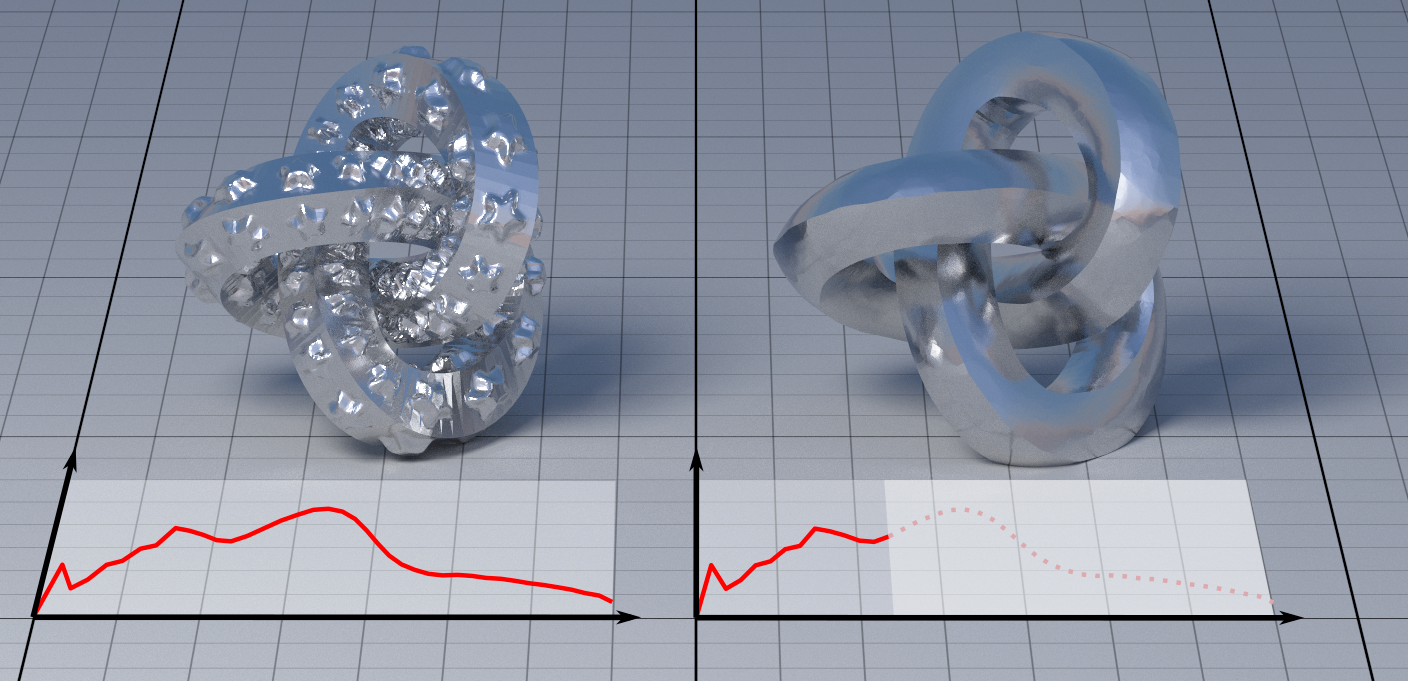}
  \end{overpic}
  \caption{\label{fig:teaser}Our method allows us to \sout{isolate and} process the geometric details of a given shape into well separated spectral bands having different semantics. Each band corresponds to a different geometric feature, possibly not localized in space. In this example, the star decorations are removed from the knot via a simple low-pass filter in the TV spectrum. The method is rotation-invariant, as it does not depend on the orientation of the knot \sout{model} in 3D space.}
  \end{center}
\end{figure}

The present work finds its motivation in the fact that detail preservation is a strict necessity in a wide class of applications in graphics. We start from the observation that, similar to most natural images, shape geometry often has a sparse gradient. For a scalar function $f$, this quantity can be measured via the {\em total variation} functional:
\begin{align}
\label{eq:tv}
    TV(f) = \int || \nabla f(x) || dx\,,
\end{align}
which, for coordinate functions, captures the amount of sharp geometric variations in a given object.
Leveraging recent progress in image processing \cite{Gilboa:2013,Burger:2016}, we use the above functional to derive a {\em nonlinear} operator, the subdifferential $\partial TV$, which has numerous piecewise-constant functions as \emph{eigenfunctions}. 

Using the TV functional, we provide a spectral decomposition ({\em analysis}) algorithm for functions $f$ on a manifold which, similarly to Fourier decompositions, carries a canonical multiscale ordering, providing a sharp separation of the source signal into different bands; as a direct consequence to the sparsity of the gradient, frequency bands in the spectral domain assume a semantic connotation, and can be filtered and processed following classical signal processing paradigms (see Figure~\ref{fig:teaser} for an example). Importantly, since $TV(f)$ remains finite for a wide variety of discontinuous $f$ (with an appropriate adaptation of the definition~\eqref{eq:tv} to be discussed in Eq.~\eqref{dfn:dualTV}), discontinuities are preserved by the transform, \sout{which is} and is therefore well-suited for representing jumps and sharp high-frequency variations. The inverse transform ({\em synthesis}) is realized by a single linear step.

Our approach shares some common traits with Laplacian mesh processing, for which it provides a nonlinear alternative; in particular, we show that both settings can be derived as specific instances of a more general model.
We demonstrate the versatility of our framework by applying it with success to several applications in geometry processing and graphics, relating its multiple incarnations to existing ad-hoc approaches, and illustrating its performance across different settings and shape representations.

\subsection{Contribution}
With this paper we introduce a new framework for geometry processing that weaves discontinuity-preserving regularization into the fabric of a spectral approach.
Our key contributions are:
\begin{itemize}
    \item We introduce a general {\em spectral} framework to analyze and process non-smooth signals defined on surfaces, as well as the surfaces themselves, which fully preserves geometric detail; 
    \item We adopt a fast, unconditionally stable algorithm to solve the resulting nonlinear problem on non-Euclidean domains, with limited dependence on a few parameters;
    \item We analyze the evolution of surfaces along the flow induced by the operator, and relate it to existing geometric flows\sout{, and further inspect the commonalities with the Laplace-Beltrami operator}\sout{verifying that they both are particular cases of a more general operator};
    \item We explore a wide range of possible applications in shape analysis and modelling, where we compare to state-of-the-art methods in the respective domains, demonstrating production quality results in many cases.
\end{itemize}
\section{Related Work}

From a general perspective, ours is a {\em filtering} approach that incorporates an edge-preserving diffusion process within a spectral decomposition framework, and draws inspiration from recent progress in nonlinear image processing. 
In the following, we cover the works that more closely relate with these themes. 

\subsection{Variational mesh processing}
Our methodology falls within the class of energy minimization methods for geometry processing, and carries a natural dual interpretation as a diffusion-based approach. 

\vspace{1ex}\noindent\textbf{Diffusion-based methods.}
Following the work of Taubin \shortcite{Taubin:1995}, classical editing tasks such as surface smoothing, denoising and enhancement have been traditionally phrased as variants of the diffusion PDE $\partial_t u = \Delta_\mathcal{M} u$. Here, $u$ is a signal defined on the surface $\mathcal{M}$ \sout{(e.g., the vertex or normal coordinates for each point)}, and $\Delta_\mathcal{M}$ is its Laplace-Beltrami operator. For example, feature-aware denoising driven by local curvature has been addressed by modifying the metric on $\mathcal{M}$ (and hence its Laplacian), leading to a rich production of anisotropic techniques \cite{DMSB-gi2000,tasdizen2002geometric,Clarenz:2002,bajaj03}. Higher-order filtering with volume preservation to prevent surface shrinking was proposed in \cite{Desbrun:1999}; the same work clarified the well known duality relating the diffusion to mean curvature flow, and showed how the latter notion leads to \sout{an alternative and} a more robust formulation for denoising applications. This was later modified in \cite{kazhdan2012can} to resolve \sout{numerical} instabilities \sout{occurring across} of the flow. 

Our approach also involves a diffusion process, and comes with an associated geometric flow. 
We identify two key differences \textcolor{red}{over} \sout{as related to} prior work: (1) in our case, the operator involved is nonlinear, and (2) diffusion takes place {\em within} the level sets of the signal $u$ rather than {\em across} the boundary of the level sets. As a consequence, discontinuities are not smeared throughout the diffusion process, while contrast is reduced. In the inset figure, the source signal (left) is 
\setlength{\columnsep}{7pt}
\setlength{\intextsep}{3pt}
\begin{wrapfigure}{r}{0.45\linewidth}
\vspace{-0.15cm}
\begin{center}
\includegraphics[width=\linewidth]{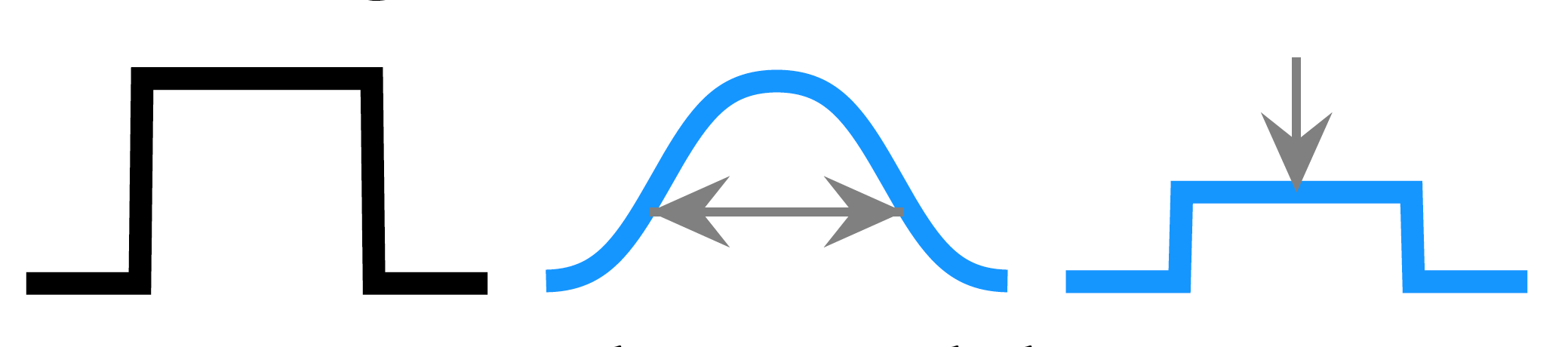}
\end{center}
\end{wrapfigure}
processed via the classical ``horizontal'' isotropic diffusion (middle) and with our ``vertical'' diffusion (right); in the latter case, the \sout{step} width remains constant while the amplitude decreases. When the signal encodes surface geometry, this implies that sharp features are preserved.

\vspace{1ex}\noindent\textbf{Energy minimization methods.}
%
Other approaches are based on formulating editing operations as the minimization of an energy. Although mathematically any variational \sout{minimization} problem can be rephrased as a diffusion problem, \sout{taking one of the two perspectives may be more or less convenient for computational reasons. Typically,} \sout{formulating a variational problem} choosing the former enables the implementation of a richer class of algorithms. 
%
%
Such methods lift ideas from \sout{the} image processing \sout{area} to the surface domain. For example, in \cite{He:2013} mesh denoising is done by minimizing the $L_0$ gradient norm of the vertex positions, thus suppressing low-amplitude details, as done in \cite{xu:2011} for images; extensions to meshes of bilateral \cite{tomasi:1998} and mean shift \cite{meanshift} filtering were proposed in  \cite{fleishman03,Solomon:2014}\sout{ respectively}; \textcolor{red}{a variant of \cite{Zhang_rolling:2014} was applied to the normal vector field in \cite{wang:2015}}; \sout{more recently,} shock filters \cite{shock90} were applied to surface geometry for feature sharpening and enhancement in \cite{Prada:2015}. Gradient-domain mesh processing \cite{Chuang:2016:GPO} is also a recent direction inspired by success in the image domain, allowing explicit prescription of target gradient fields.\sout{ Other filtering methods allow prescribing vertex positions \cite{kobbelt98,Sorkine:2005} or the Riemannian metric directly \cite{chuang:2011}.} 

The above testifies to a wealthy literature of nonlinear edge preserving strategies for mesh filtering. The main drawback of these methods lies in their lack of generality. Here we provide a general framework to perform multiscale analysis and synthesis of arbitrary signals defined on geometric domains (meshes and point clouds alike); while denoising, fairing and enhancement can be easily realized within this framework, we are not limited to these applications and provide a more flexible setting for geometry processing.

\vspace{1ex}\noindent\textbf{TV-based methods.}
\sout{Ours is certainly not the first work to consider the total variation functional in a geometry processing context.} TV and the strictly related Mumford-Shah functional have been considered before for different tasks in geometry processing, \textcolor{red}{e.g. surface reconstruction \cite{Liu:2017}}. 
 Zhang et al.~\shortcite{Zhang:2015} consider an extension of the ROF model \cite{Rudin:1992} for denoising surfaces, which they solve with an augmented Lagrangian method followed by a geometry reconstruction step \cite{Sun:2007}. Zhong et al.~\shortcite{Zhong:2018} follow a similar approach, with the inclusion of an additional regularizer defined as the Laplacian energy of the normal coordinates, which helps dampen the staircase effect that is observed in \cite{Zhang:2015}.
\sout{The resulting energy is also minimized as in Zhang 2015.}
These methods are related to ours, \sout{since} as they \sout{both} involve the minimization of the TV of the surface normal field; we also consider this kind of energy, but in addition, we study the {\em geometric flow} associated to it. Further, our approach is not restricted to normal fields, but allows processing arbitrary surface signals\sout{ defined on the surface}. Tong and Tai~\shortcite{Tong:2016} consider the Mumford-Shah functional \sout{, instead of the TV functional,} as a measure of sharp geometric variation for a given signal. Their focus is on the detection of features lines (i.e. the boundaries between neighboring segments) for scalar functions such as mean curvature and color texture, rather than geometry processing. The \sout{Mumford-Shah} MS functional has also been considered in \cite{Bonneel:2018} in its Ambrosio-Tortorelli approximation. The authors extend the range of possible applications of \cite{Tong:2016} by operating on the surface normal field, showing results in  inpainting, denoising and segmentation tasks. In addition to lacking an analysis of the TV normal flow, none of these methods provide a {\em spectral} framework based on the TV functional for geometry processing.
\subsection{Spectral geometry processing}
%
As shown in Figure~\ref{fig:teaser}, our multiscale approach provides a sharp separation of an input signal into ``feature bands''. 
We achieve this by means of a generalized {\em spectral} decomposition. Spectral geometry processing was introduced \sout{again} by \sout{the work of} Taubin~\shortcite{Taubin:1995} and further developed in \cite{Sorkine:2005,Levy:2006,rong2008spectral,Zhang:2010} for tasks of shape reconstruction, modelling, and deformation transfer among others. These works generalize Fourier analysis to 
\setlength{\columnsep}{7pt}
\setlength{\intextsep}{1pt}
\begin{wrapfigure}[6]{r}{0.55\linewidth}
\vspace{-0.15cm}
\begin{center}
\begin{overpic}
[trim=0cm 0cm 0cm 0cm,clip,width=1.0\linewidth]{./images/gibbs2.png}
\vspace{0.1cm}
\put(2.5,-20.8){\frame{\includegraphics[width=0.25\linewidth]{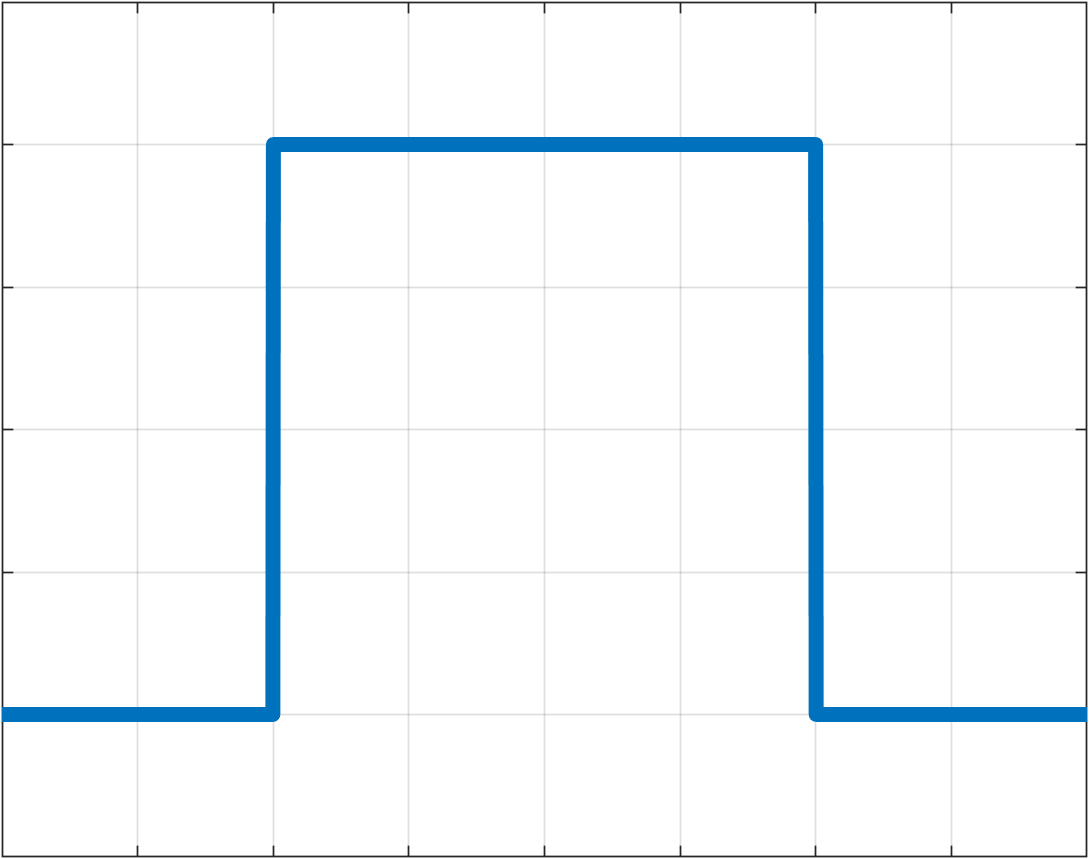}}}
\put(33.5,-20.8){\frame{\includegraphics[width=0.25\linewidth]{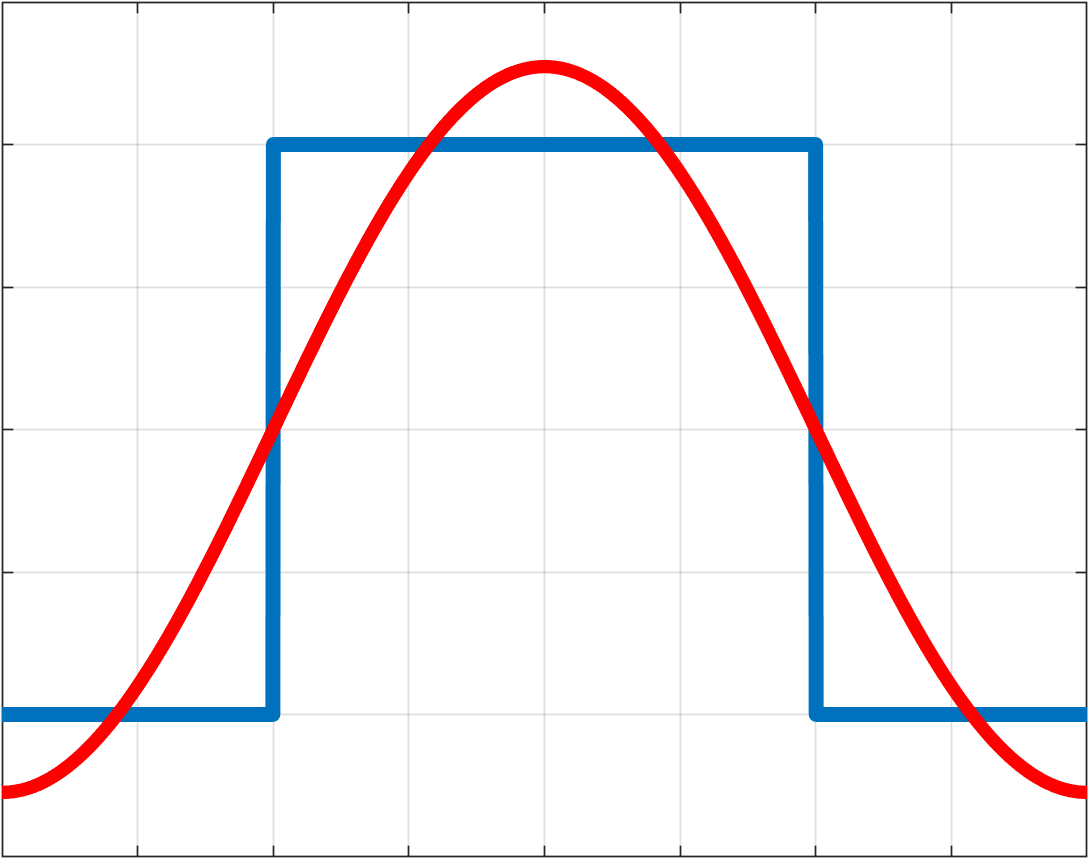}}}
\put(64.5,-20.8){\frame{\includegraphics[width=0.25\linewidth]{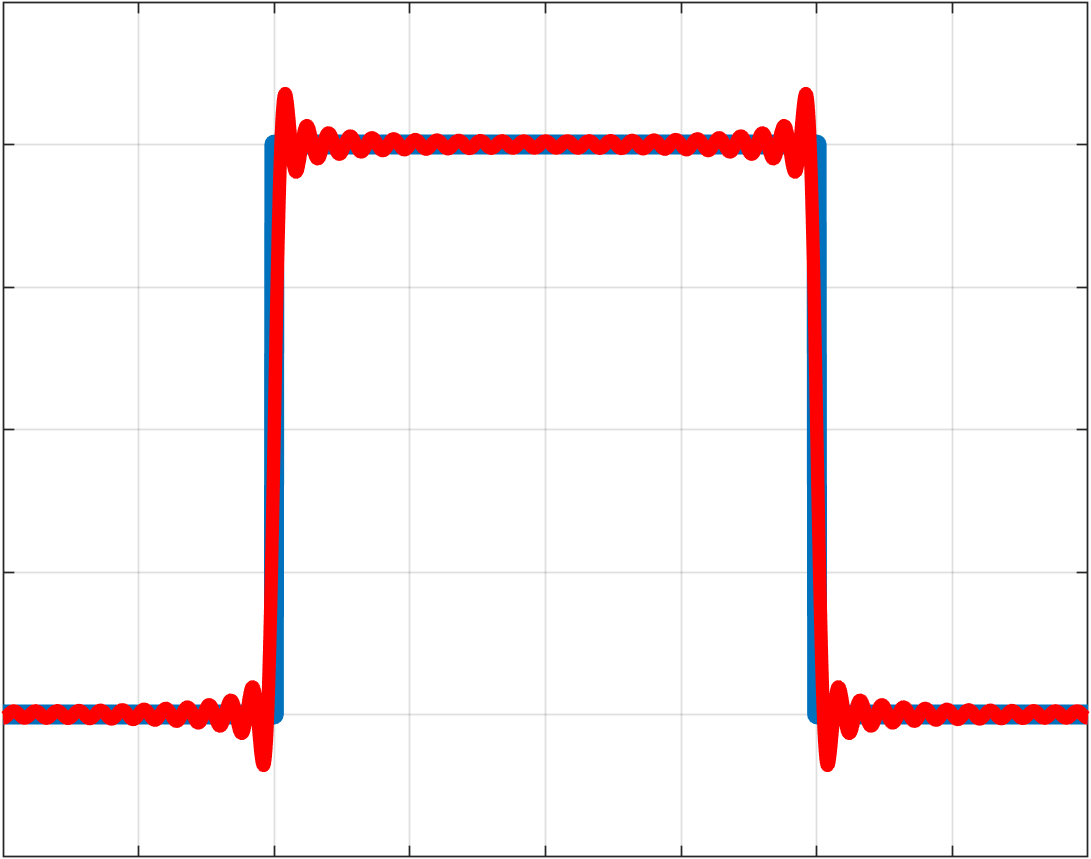}}}
\end{overpic}
\end{center}
\end{wrapfigure}
%
  surfaces by observing that the Laplacian eigenvectors form an orthogonal set of smooth basis functions with a canonical ordering (low to high frequencies). This basis is optimal for representing smooth functions compactly \cite{Aflalo:2014}, but it gives rise to \sout{undesirable} artifacts in the presence of sharp variations, thus \sout{greatly} reducing the benefits of working with a spectral representation; in the inset, we show the typical blurring effect (middle) and Gibbs oscillations (right) due to the \sout{presence of} jumps in the source signal (left), \sout{which are} not handled well by the smooth Laplacian eigenbasis.

Variants of the Laplace operator \sout{have been considered in order to} add flexibility and \sout{incorporate} feature awareness\sout{ in the spectral basis}, including anisotropic \cite{Andreux:2015} and extrinsic \cite{Liu:2017} variants, see \cite{Wang:2019} for a recent survey. Sparsity \sout{in the context of spectral geometry} has \sout{also} been analyzed in \cite{Neumann:2014}, where Laplacian eigenfunctions are modified to have limited support by penalizing their $L^1$ norm. We differ in that we consider signals with sparse gradients instead.  
 \sout{Despite recent progress,} All these operators are linear, and their associated energies promote smoothness\sout{ (we formalize this statement more precisely below)}. Our operator is nonlinear, hence more powerful, and its \sout{(generalized)} eigenfunctions are well suited to represent jump discontinuities in the signal. This yields a cleaner separation of different features into different bands, \sout{; this is} in contrast to Laplacian-based representations, which spread individual features across the entire spectrum.




%

\subsection{Nonlinear image processing}
%
%
Finally, similar to prior research, our approach is also based on ideas originating from image processing.
In this field, nonlinear variational methods have replaced linear methods due to their higher precision in approximating natural phenomena. To date, a major role has been played by the total variation (TV) energy and its minimization, originally exploited by Rudin et al.~\shortcite{Rudin:1992} with the ROF model for image denoising. Other uses of the TV functional span image deblurring, inpainting, interpolation, image decomposition, super-resolution and stereovision, to name but a few; for a general introduction to the topic see \cite{chambolle:2010}.
Recently, a spectral theory related to the gradient flow of this functional (c.f. Bellettini et al.~\shortcite{bellettini:2002}) on eigenfunctions, was developed by Gilboa \shortcite{Gilboa:2013,GuySpectralSiam14} and generalized in \cite{Gilboa:2015, Burger:2016} with applications to image processing, allowing for the analysis of images at different scales by exploiting the edge-preserving property of the total variation. Here we propose to lift this latter framework from the Euclidean domain to surfaces, put it in relation to existing geometry processing models, and demonstrate several possible applications in the area of 3D graphics.

\section{Mathematical Preliminaries}


\subsection{Surfaces}
We model shapes as compact and connected Riemannian manifolds $\M$ embedded in $\mathbb{R}^3$, possibly with boundary $\partial\M$, and with tangent bundle $\mathcal{T}\M =\bigcup_{p \in \M} \mathcal{T}_{p} \M$. The manifold is equipped with a metric, i.e., an inner product defined locally at each point, $\langle \cdot,\cdot\rangle_p: \mathcal{T}_{p}\M \times \mathcal{T}_{p}\M \rightarrow \mathbb{R}$, making it possible to compute lengths \sout{, areas,} and integrals on $\M$. 
 The intrinsic gradient $\nabla_{\M}$, defined in terms of the metric, generalizes the \sout{classical} notion of gradient to manifolds; 
%
 similarly, the intrinsic divergence $\mathrm{div}_\M$ \sout{generalizes the notion of divergence and} can be defined as the negative adjoint of the gradient, assuming Neumann boundary conditions:
 %
%
 %
\begin{align}
&\int_{\M}  \mathrm{div}_{\M}V f  \  dx  = - \int_{\M} \langle V,\nabla_\M f \rangle_p \  dx\,,
\end{align}
where $f:\M\to\mathbb{R}$ is a scalar function and $V:\M\to\mathcal{T}\M$ is a vector field tangent to the surface.
In the following, to simplify the notation we will drop the subscript $\M$ whenever clear from the context.
We further denote by $\mathcal{L}^p(\M)=\{ f:\M\to\mathbb{R} ~| \int_\M |f(x)|^p dx <\infty\,, p\ge 1 \}$ the function space of $p$-integrable functions on $\M$.

\subsection{Total Variation}
%
 The total variation of a differentiable function $f$ on $\M$ is defined as
\begin{align} \label{mth:gradtv}
TV(f)=\int_{\M} \| \nabla f(x) \| dx  \,,
\end{align} 
where $\|\cdot\|$ is the norm induced by the inner product $\langle \cdot,\cdot\rangle_p$.
\sout{The total variation is a} \textcolor{red}{This} functional \sout{that} quantifies the amount of oscillations in a given signal. To account for non-differentiability of $f$, we adopt a weaker definition of TV  
 in terms of continuously differentiable test vector fields $U:\M\to\mathcal{T}\M$ \cite{benartzi:2006}: 
%
%
\begin{align} \label{dfn:dualTV}
    TV(f)=\sup_U\left\{ \int_ {\M} f(x) ~ \mathrm{div} U(x) \,  dx :  ~ \|U(x)\|\leq 1, ~ \forall x \right\} \,.
\end{align}
%
Alternatively, one can define an \emph{anisotropic} variant of the standard TV by considering $U(x)$ to be bounded in the $L^\infty$ norm. We refer to functions $f$ of bounded variation, i.e. $TV(f)< +\infty$, to be in $BV(\M)$.

 
The TV functional is related to  fundamental geometric notions such as the curvature of level sets of a function, through the celebrated co-area formula \cite{Fleming:1960}. One particular case of this relation concerns the total variation of indicator functions of closed sets; in this case, the TV corresponds to the perimeter of the set (see \cite[Sec. 1.3]{chambolle:2010} for the Euclidean setting), which generalizes to manifolds in a straight forward way:
\begin{thm} \label{th:perimeter}
Let E $\subset \M$ be a measurable closed subset of $\M$, with smooth boundary $\partial E$, and let $\chi_E$ be its indicator function. Then:
\begin{align}
    TV(\chi_E)= \mathrm{length}(\partial E)\,.
\end{align}
\end{thm}
\noindent{\em Proof}: See Appendix \ref{apx:tper}.

\vspace{1ex}
As we show next, this particular class of indicator functions includes the set of generalized eigenfunctions of the operator associated to the TV functional.

\section{Spectral TV decomposition}
%
%
%

In this work we introduce a new invertible transform for signals defined on surfaces, with the following key properties:
\begin{itemize}
    \item The forward transform is non-linear;
    \item The inverse transform is linear;
    \item Any signal $f\in BV(\M)$ on the surface can be decomposed and reconstructed up to a desired accuracy;
    \item Discontinuities in $f$ are preserved by the transform.
\end{itemize}
The following considerations closely follow the ideas of \cite{Gilboa:2013, GuySpectralSiam14} in the Euclidean setting, but extend them to functions on manifolds to enable discontinuity-preserving geometry processing.

\subsection{TV flow}
Consider the following energy for surface signals $u,u_0\in BV(\M)$:
\begin{equation} \label{eq:energy}
E(u) = \frac{1}{2}\int_\mathcal{M}(u(x)-u_0(x))^2 dx+ t TV(u)\,, \quad t>0
\end{equation}
with the associated variational problem: 
\begin{equation}\label{eq:1}
\min_{u\in BV(\M)} E(u)\,. 
\end{equation}

The above seeks to minimize a functional over two terms: a \textit{fidelity term}, which quantifies the distance from an initial function $u_0$, and a \textit{regularization term} which measures the total variation of the sought minimizer. The regularization parameter $t$ balances between the two terms. 
In image processing, where $\mathcal{M} = S \subset \mathbb{R}^2$, Eq.~\eqref{eq:1} is known as the ROF model for image denoising \cite{Rudin:1992}, and dissipates noise in flat regions of an image $u_0$ while preserving object contours.

\begin{figure}[t]
  \centering
\begin{overpic}[trim=0cm 0cm 0cm 0cm,clip,width=1.0\linewidth]{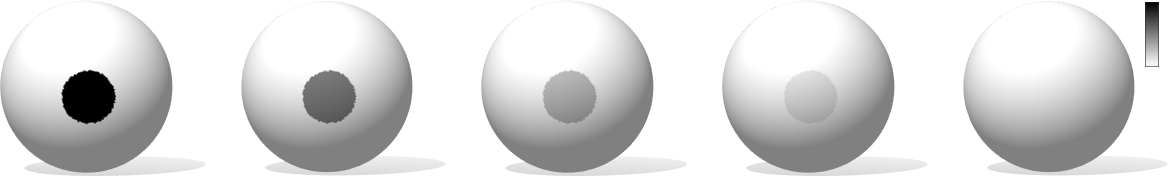}
\put(4,-3){\footnotesize $t=0$}
\put(23,-3){\footnotesize $t=0.05$}
\put(44.5,-3){\footnotesize $t=0.1$}
\put(65,-3){\footnotesize $t=0.2$}
\put(85.5,-3){\footnotesize $t=0.3$}
\put(99.7,9.5){\tiny $0$}
\put(99.7,13.5){\tiny $1$}
\end{overpic}
  \caption{\label{fig:TVflowsphere}Evolution of a disk-shaped function along the TV flow on a spherical surface. Since the diffusion only acts {\em inside} the level sets, the disk preserves its shape and loses contrast linearly until it vanishes.
  }
\end{figure}
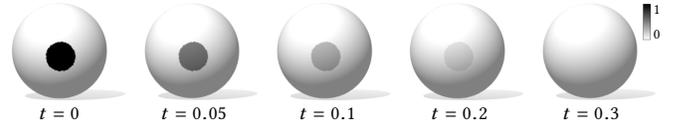

Iteratively applying Eq.~\eqref{eq:1} to the output of a previous solution with small increases of $t$ can be seen as a discretization of a time continuous \textit{total variation flow}: 
%
\begin{align}\label{eq:pde}
    \begin{cases}
    \frac{\partial u}{\partial t}= ~\mathrm{div} \left(\frac{\nabla u}{\|\nabla u\|}\right) &  \text{in  } \M,\\
    \langle \nabla u , \vec{n} \rangle=0 & \text{on  }  \partial \M , \\ 
    u(0,x) = u_0,
    \end{cases}
 \end{align}
where $\vec{n}$ is the unit vector normal to the boundary $\partial\M$. Equations~\eqref{eq:pde} describe a physical process of isotropic diffusion within the level sets of the initial function $u_0$, without any diffusion across them (see Figure~\ref{fig:TVflowsphere}). In $\mathbb{R}^n$, it is guaranteed that the flow converges in finite time to a constant solution under mild assumptions \cite[Cor.~1]{andreu:2002}; no proofs exist for manifolds, but in our experiments we always observed convergence.

To give sense to Eq.~\eqref{eq:pde} for nondifferentiable $u$ and points at which $\nabla u(x)=0$, one needs to turn to the \textit{subdifferential}, which for a convex functional $J$ (TV in our case) over a function space $\mathcal{X}$ is formally defined as the set:
\begin{align*}
    {\partial J}(u) =
    \left\{ p \in \mathcal{X}^* ~|~  \ J(v) - J(u) \geq  \langle p,v-u\rangle, \ \forall v \in \mathcal{X} \right\}\,,
\end{align*}  
with $\mathcal{X}^*$ being the dual space of $\mathcal{X}$. For smooth $u$ with non-zero gradient one obtains:
%
\begin{align}\label{eq:dtv}
    \partial TV (u) = \left\{ -\mathrm{div} \left(\frac{\nabla u}{\|\nabla u\|}\right)\right\}\,.
\end{align}
%
We refer the reader to \cite{bellettini:2002} for more details on the TV flow in a Euclidean setting.

%
%

%
%

\subsection{Nonlinear eigenvalue problem}

A generalized eigenfunction of the operator $\partial J$ is a function $u$ with $\|u\|=1$ satisfying the following relation for some number $\lambda\in\mathbb{R}$ (its corresponding generalized eigenvalue):
\begin{align}\label{eq:eigentv}
    \partial J(u) \ni \lambda u\,,
\end{align} 
where $J=TV$ in our case. 
%
This relation generalizes the classical characterization of eigenpairs of a linear operator: 

\begin{remark}
For the Dirichlet functional $D(u) = \frac{1}{2}\int_\M \| \nabla u \|^2$, one obtains $\partial D(u) = \{-\Delta u\}$, such that the analogy to Eq.~\eqref{eq:pde}  becomes the standard diffusion equation, and Eq.~\eqref{eq:eigentv} reduces to the standard eigenvalue problem $\Delta u = \lambda u$ (up to a sign). 
\end{remark}



\subsection{The TV flow on eigenfunctions}
\label{section:4:3}
Let us recall the ideas from \cite{Gilboa:2013, Gilboa:2015} to motivate a spectral decomposition using the TV flow. For eigenfunctions $u_0:\M\to\mathbb{R}$ with eigenvalue $\lambda$, we get a simple analytic solution to the flow of Eq.~\eqref{eq:pde}:
\begin{align*} 
    u(t,x)=\begin{cases}
    (1 - t \lambda) \, u_0(x) & \text{for   } t \leq \frac{1}{\lambda} \\
    0  & \text{otherwise}
    \end{cases}
\end{align*}
At each point, the eigenfunction decreases linearly at increasing $t$, until it goes to zero. In other words, TV eigenfunctions preserve their shape (up to rescaling) under the action of the flow; Figure~\ref{fig:TVflowsphere} illustrates one such eigenfunction. 

\begin{figure}
\centering%
\includegraphics[width=\linewidth]{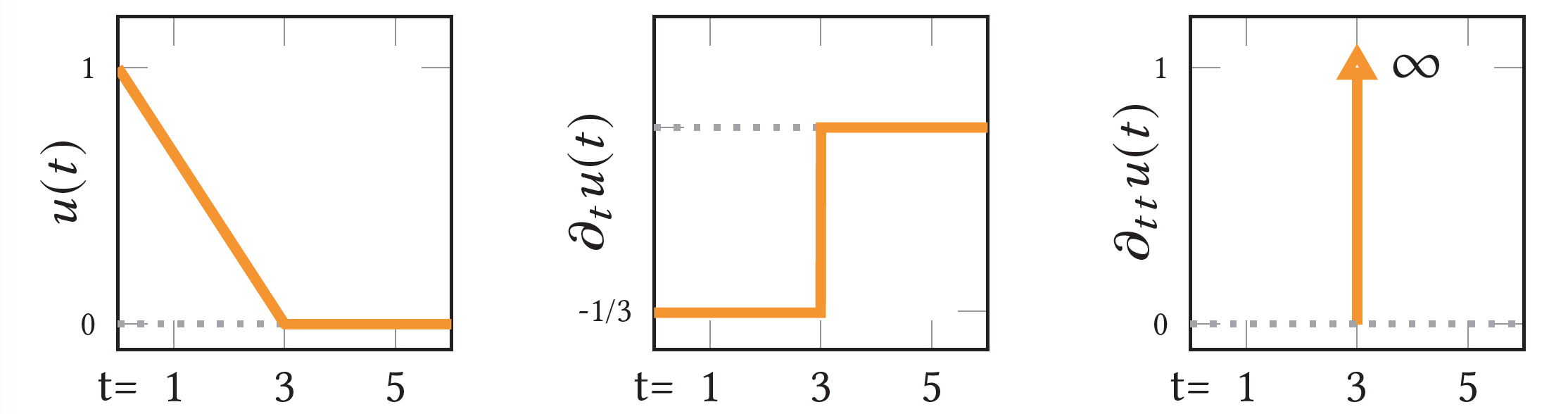}
\caption{\label{fig:sol1dflow}Analytic solution (left) to the 1D flow for a TV eigenfunction $u_0$ with eigenvalue $\lambda=\frac{1}{3}$, first and second derivatives (resp. middle and right).}%
\end{figure}

The first derivative of $u(t,x)$ w.r.t. time is \sout{the} piecewise-constant\sout{ function}:
\begin{align*} 
    \partial_t u(t,x)=\begin{cases}
     - \lambda u_0(x) & \text{for   } t \leq \frac{1}{\lambda} \\
    0  & \text{otherwise}
    \end{cases}
\end{align*}
Considering also the second derivative (in a distributional sense), we get a delta function placed exactly at the point in time where the solution $ u(t,x)$ vanishes (see Figure~\ref{fig:sol1dflow}):
\begin{align*} 
    \partial_{tt} u(t,x)=\begin{cases}
    + \infty & \text{for   } t = \frac{1}{\lambda} \\
    0  & \text{otherwise}
    \end{cases}
\end{align*}
\sout{We see that} The behavior of the second derivative $\partial_{tt} u(t)$ for an eigenfunction $u_0$ resembles \sout{bears a strict resemblance with} the notion of spectrum in \sout{classical} Fourier analysis, since sinusoidal functions (eigenfunctions of the Laplace operator) correspond to \sout{are represented as} deltas \sout{functions} in the frequency domain. 
This motivates the definition:
\begin{defn}
\label{def:spectralRepresentation}
We refer to the distribution over time $\phi_t= \partial_{tt} u(t)$ as the {\em spectral representation} of $u_0$, and to each individual distribution at the time scale $t$ as a {\em spectral component}. 
\end{defn}
As we have seen, for a TV eigenfunction with eigenvalue $\lambda$, the spectral representation has a single component: \sout{which is} a delta at $t=\frac{1}{\lambda}$. 
However, not every spectral component $\phi_t$ is guaranteed to identify a TV eigenfunction. 
Further, at any given time scale $t$, the spectral component $\phi_t:\M\to\mathbb{R}$ still has a spatial extent over $\M$. 
To show the entire spectral activity across different time scales, we can integrate each $\phi_t$ and simply plot the function: 
\begin{align}
\label{eq:spectrum}
s(t)=\int_\M |\phi_t(x)| dx\,.
\end{align}
%
Looking at the behavior of the function $s(t)$ can be rather useful in practice, since it helps in defining the filtering operations required by the applications; examples of $s(t)$ are shown in Figure~\ref{fig:teaser}.

%
The spectral representation can be translated back to the primal domain by the linear operation of integrating over time (i.e., across all spectral components $\phi_t$) and adding the mean of $u_0$: 
\begin{align}\label{eq:inv}
 u_0= \int_0^\infty t \phi_t dt \, + \text{mean}(u_0)\,.
\end{align}
We refer \sout{the reader} to \cite{Gilboa:2013, GuySpectralSiam14} for more details (with the slight difference that we decided to define the spectral decomposition without the factor $t$ and include it in the reconstruction instead). 


\subsection{The spectral representation induced by the TV flow}\label{sec:dict}
In light of the previous discussion that TV eigenfunctions are the fundamental atoms of the spectral representation of Def. \ref{def:spectralRepresentation} (in the sense that they yield peaks in the spectral domain), one might expect to represent arbitrary signals as linear combinations of such atoms. 
Although there seems to be a vast number of TV eigenfunctions to allow such a representation\footnote{See e.g. \cite{Steidl:2004} for a proof that the Haar basis is a proper subset of the TV eigenfunctions on the real line, or \cite{Alter05} for a characterization of sets whose characteristic function is a TV eigenfunction.}, computing a \sout{reasonable (i.e.,} \textit{sparse} \sout{)} representation turns out to be difficult since, e.g., a Rayleigh principle to compute orthogonal eigenfunctions does not hold \cite{Benning:2013}.

Interestingly, Def.~\ref{def:spectralRepresentation} allows to show that the gradient flow with respect to a regularization with suitable properties does yield such a decomposition with the $\phi_t$ representing differences of eigenfunctions, see \cite{Burger:2016}. 
 While the TV in Euclidean spaces of dimension larger than $1$ does not satisfy the required properties, the spectral representation $\phi_t$ has been shown to still yield a meaningful, edge-preserving, data-dependent multi-scale representation in the field of image processing \cite{GuySpectralSiam14, Gilboa:2015}, which is why we utilize it on manifolds in a similar fashion.

\subsection{Computing the TV spectral decomposition}
\label{section:4:5}
%
To compute the spectral representation $\phi_t$ of a given function $f$, we use an implicit Euler discretization of the TV flow: We compute a new iterate by solving \sout{the ROF model of} Eq.~\eqref{eq:1} using the previous iterate in the data fidelity term starting from $u_0 \equiv f$. 
%
%
The procedure is summarized in Algorithm~\ref{alg:1}.

\begin{algorithm}
\SetAlgoLined
\SetKwInOut{Input}{input}\SetKwInOut{Output}{output}
\Input{signal $u_0$, step length $\alpha$, no. of time steps $N$}
\Output{spectral representation $\phi_t$}
\Begin{
	\For{$t=0:N$}{
		$u(t) \xleftarrow{Alg. \ref{PDHG}} \min_u TV(u) + \frac{1}{2 \alpha} || u- u_0||_2^2 $\\
		output $\phi_t \xleftarrow{} \partial_{tt} u(t)$ \\
		increase $\alpha$ \\
		update $u_0$ ${\xleftarrow{} u(t)}$
			}
\caption{\label{alg:1}Spectral TV decomposition}}
\end{algorithm}
The algorithm simply evolves the input signal by $N$ discrete steps along the TV flow; each iteration moves a step forward, with diffusion time equal to $\alpha$. As changes happen quickly for small $t$ and tend to become slower for larger $t$ we iteratively increase the step size $\alpha$ of the evolution. Subsequently, the spectral representation $\phi_t$ is constructed incrementally using finite differences, such that the integral of Eq.~\eqref{eq:inv} becomes a simple weighted sum over the $\phi_t$. We give selection strategies for $\alpha$ and $N$ in Appendix~\ref{sec:params}.

\begin{figure}
    \centering
     \includegraphics[width=0.9\linewidth,trim={0 0 0 0},clip]{images/duck.png}
      \caption{\label{fig:duck}Spectral TV transform \sout{of a color signal} on a rubber ducky. The original signal (top left) is composed of three indicator functions of different regions, \sout{which are} represented in the TV spectral domain as three separated peaks (top right). Each plot visualizes the spectral activity \sout{$s(t)$} of the corresponding surface signal\sout{, defined as  $s(t)=\int_\M |\phi_t(x)| dx$}. By filtering the peaks in the spectral domain we can single out each component at a different scale, obtaining one region at a time (bottom row).}
      
\end{figure}
\begin{figure}[b!]
    \centering
\begin{overpic}[width=\linewidth]{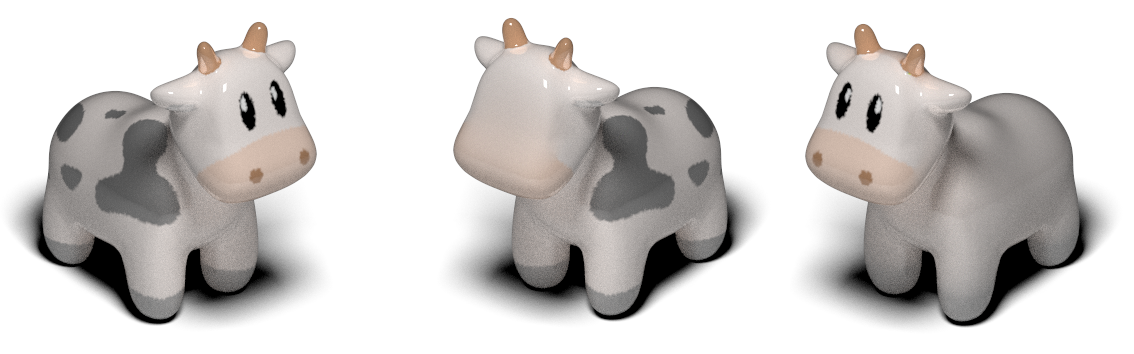}
\put(10.5,-1){\footnotesize source}
\put(48,-1){\footnotesize filtered A}
\put(80,-1){\footnotesize filtered B}
\end{overpic}
\caption{\label{fig:cow}The source surface signal (left) is transformed into a spectral TV representation, and processed with two band-pass filters to obtain a cow without eyes and nostrils (A) and without spots (B).}
\end{figure}

\begin{remark}
This yields a {\em multiscale} representation for the input, since each diffusion time $t$ captures a feature at a different scale -- namely, the feature that vanishes at $t$. Here, time scales play the same role as a wavelength in classical Fourier analysis. 
\end{remark}

See Figures~\ref{fig:duck} and \ref{fig:cow} for examples of multiscale TV decomposition of color signals on two different surfaces.

%
%
    

%
\vspace{1ex}\noindent\textbf{Algorithmic considerations}. Algorithm~\ref{alg:1} is unconditionally stable, and line 3 corresponds to taking an implicit Euler step along the discretized TV flow. This property allows us to use arbitrary values for the step size $\alpha$ (therefore, the resolution of the decomposition) without compromising convergence. 
 In practice, in our experiments we use a variant of Algorithm~\ref{alg:1} and study the evolution of the flow backwards in time by following a nonlinear inverse scale approach,
 that starts from a constant solution and converges to the initial function $u_0$. The spectral decompositions of both approaches are very similar in practice and have been proven to be identical under certain assumptions, see \cite{Burger:2016}. The inverse scale variant comes with additional numerical robustness. \sout{, as remarked in the Appendix.} 
The complete algorithm, along with a detailed discussion on the two flows and on the role of $\alpha$ are given in Appendices~\ref{apx:inverse} and~\ref{apx:inversecomp}.
%


\subsection{Properties of TV eigenfunctions}\label{sec:props}
We conclude by giving a more complete characterization of the TV eigenfunctions on surfaces. 
%
%
On the image domain, Bellettini et al.~\shortcite{bellettini:2002} proved that any indicator function $\chi_S :\mathbb{R}^2 \to \{0,1\}$ of a {\em convex} closed subset $S \subset \mathbb{R}^2$ is a TV eigenfunction with eigenvalue $\lambda=\frac{Per(S)}{|S|}$ if it satisfies the \sout{following} isoperimetric \textcolor{red}{inequality}\sout{requirement}:
\begin{align*}\label{eq:ratio}
     \max_{q  \in \partial S} {k}(q) \leq \frac{Per(S)}{|S|}\,,
\end{align*}
\sout{This inequality states} \textcolor{red}{requiring} \sout{that} the maximal curvature $k$ of \sout{the boundary of $S$} $\partial S$ \sout{must} \textcolor{red}{to} be less than or equal to the perimeter-area ratio of $S$, intuitively \sout{, this excludes} \textcolor{red}{excluding} oblong \sout{or excessively thin} regions.

Such results have later been generalized to $\mathbb{R}^n$ as well as to unions of such convex sets, where the convex sets need to be sufficiently far apart for the TV spectral decomposition to yield a single $\delta$-peak for each of the subsets, see \cite{Alter05}.
%
%
\setlength{\columnsep}{7pt}
\setlength{\intextsep}{1pt}
\begin{wrapfigure}[3]{r}{0.42\linewidth}
\begin{center}
\begin{overpic}
[trim=0cm 0cm 0cm 0cm,clip,width=1.0\linewidth]{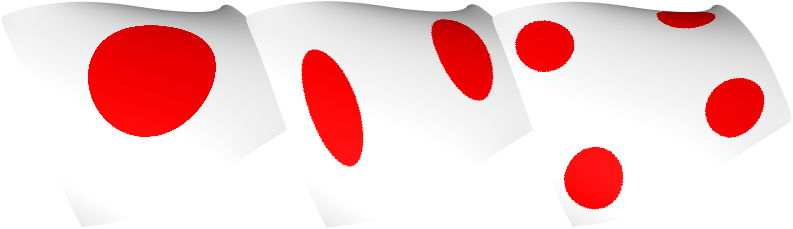}
\end{overpic}
\end{center}
\end{wrapfigure}
We conjecture (and consistently confirmed in all our experiments) that this result can be generalized to surfaces by replacing \sout{the convexity requirement} \textcolor{red}{convex subsets} with \sout{the notion of} {\em geodesically convex} subsets of surface $\M$, making the \sout{three} examples in the inset figure eigenfunctions of the \sout{total variation} \textcolor{red}{TV} on a manifold.
A patch $C \subset \M$ is geodesically convex if, given any two points $p,q \in C$, there exists a unique minimum-length geodesic \sout{$\gamma:[p,q] \to \mathcal{M}$} 
 connecting them which lies entirely in $C$. 
 \sout{This means that} \textcolor{red}{Therefore} in certain practical cases (e.g., the cow's nostrils in Figure~\ref{fig:cow}) one actually gets TV eigenfunctions in the spectral decomposition\sout{ of the given signal}.
 
%

\begin{figure}
    \centering
\begin{overpic}[width=0.9\linewidth]{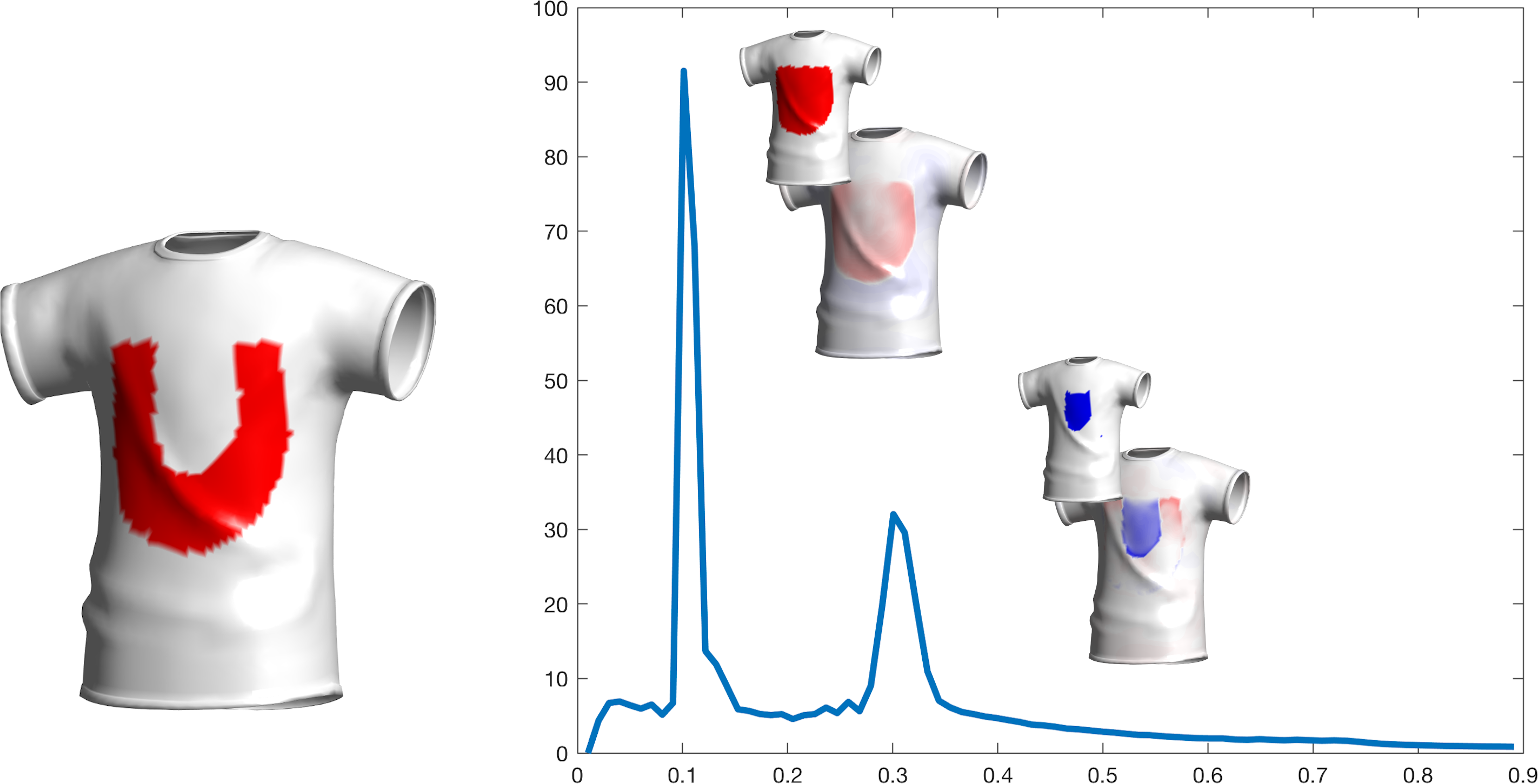}
\put(10.5,-1){\footnotesize Input}
\end{overpic}
\caption{\label{fig:shirt}
A U-shaped function (left) is decomposed as a linear combination of TV eigenfunctions corresponding to two geodesically convex sets, \sout{which are} manifested as two peaks in the TV spectrum. Functions are color-coded, growing from blue (negative values) to red (positive values).}
\end{figure}

\begin{figure}[hb!]
    \centering
\begin{overpic}[width=0.87\linewidth]{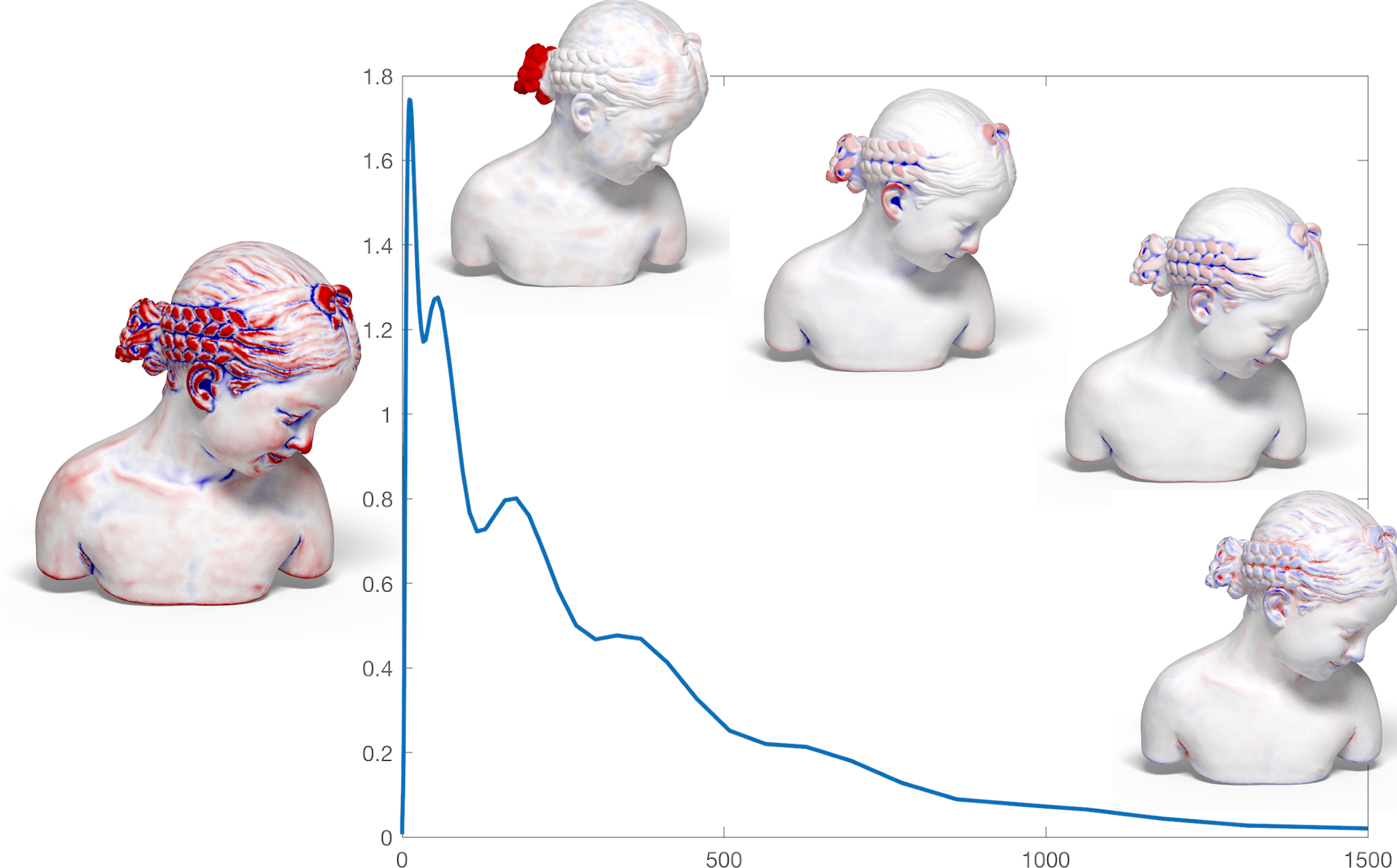}
\put(10,13){\footnotesize Input}
\put(39,39){\footnotesize $\phi_{27}$}
\put(61.5, 33){\footnotesize $\phi_{38}$}
\put(82,25){\footnotesize $\phi_{46}$}
\put(88,4){\footnotesize $\phi_{60}$}
\end{overpic}
\caption{\label{fig:spectral_bimbe}Spectral TV decomposition of the mean curvature function on the bimba (left). The \sout{resulting} spectral components $\phi_t$ are visualized at increasing index. The curve plot shows the spectral activity function $s(t)$ of Eq.~\eqref{eq:spectrum}. Low frequency components localize on semantically relevant regions (the chignon), while higher frequencies localize along feature lines (the hair strands).}
\end{figure}
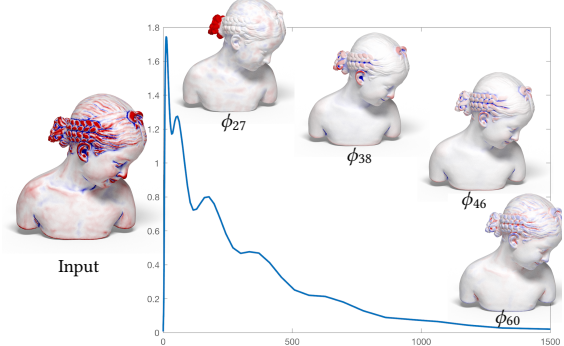

 In Figure~\ref{fig:shirt} we show an example of indicator function of a {\em non}-geodesically convex set (U-shaped, left) that can be represented by a linear combination of indicator functions of geodesically convex sets (right). In Figure~\ref{fig:spectral_bimbe} we show the spectral components of a more complex signal (the mean curvature); in this case, the components are not geodesically convex but we still get a meaningful decomposition that can be used for mesh processing.

\section{Implementation}\label{sec:impl}
%

We implemented the framework in Matlab 2019a, and ran all the experiments on an Intel i7-4558U  CPU with 16 GB RAM. \sout{The code will be made publicly available upon acceptance.}
A discussion on runtime performance \sout{together with a runtime table for our experiments are} is provided in Appendix~\ref{sec:runtime}.


\subsection{Discretization}

\begin{wrapfigure}[4]{i}{0.21\linewidth}
\centering
    \begin{overpic}[scale=0.12]{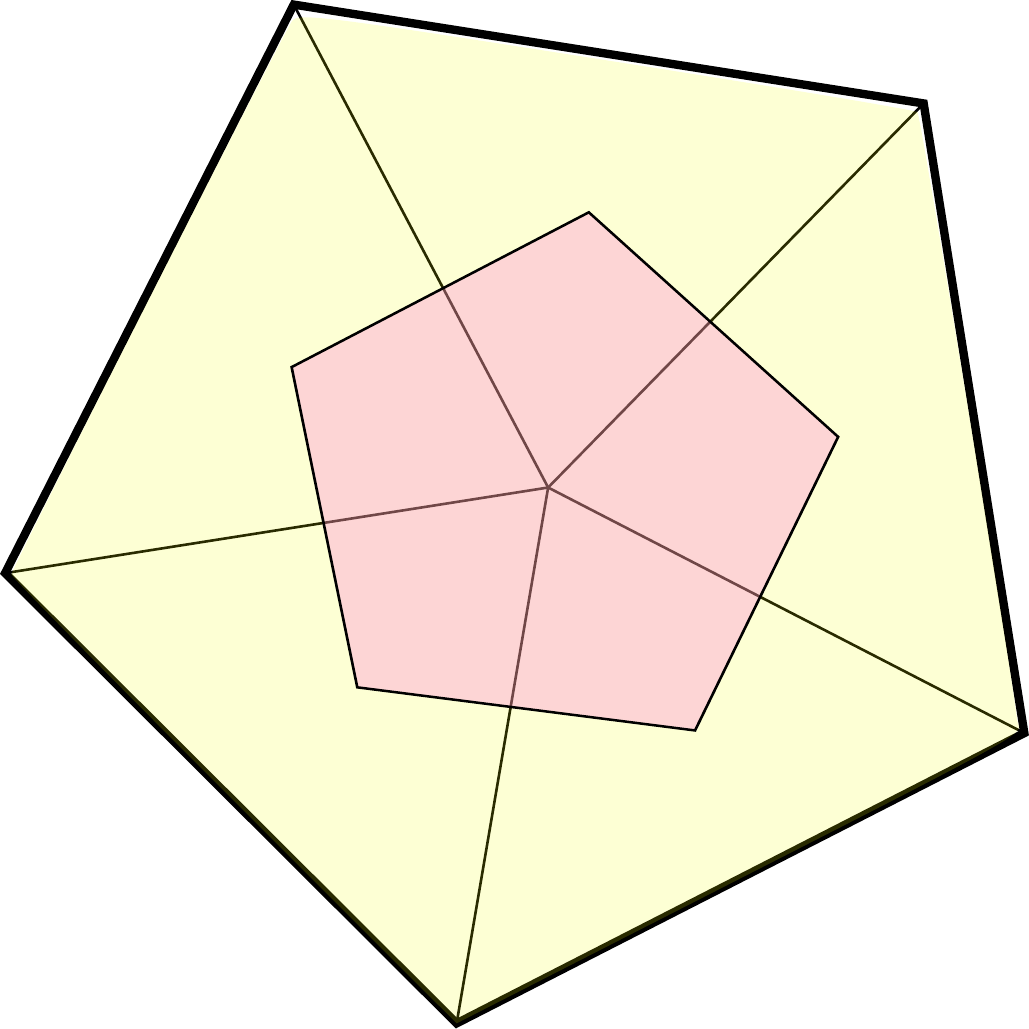}

\put(37,-15){$\mathbf{v}_j$}
\put(-30,45){$\mathbf{v}_k$}
\put(56,59){$\mathbf{v}_i$}
\put(14,57){$e_{ik}$}
\put(51,21){$e_{ij}$}
\end{overpic}
\end{wrapfigure}
\sout{In the discrete setting, we approximate} \textcolor{red}{We discretize} shapes as manifold triangle meshes $(V,E,F)$, sampled at vertices $\mathbf{v}_i, i=\{1...|V|\}$, with the constraint that each pair of triangles $(\mathbf{t}_i,\mathbf{t}_j) \in F$ shares at most one edge $ e \in E$. Functions $f: \M \rightarrow \mathbb{R}$ are approximated at vertices with piecewise-linear basis elements; tangent vector fields $U:\M\to\mathcal{T}\M$ are constant within each triangle.

The gradient operator $\mathbf{G}\in\mathbb{R}^{3 |F| \times |V|}$ takes a function value per vertex $\mathbf{v}_i$ and returns a 3D vector per triangle $\mathbf{t}_{ijk} \in F$: 
 \sout{It is computed as:}
\begin{align*}
    \mathbf{G} f(\mathbf{t}_{ijk})\hspace{-0.05cm}=\hspace{-0.05cm} 
    \begin{pmatrix}\mathbf{v}_{j}^\top-\mathbf{v}_i^\top \\ \mathbf{v}_k^\top-\mathbf{v}_i^\top\end{pmatrix}^\top \hspace{-0.1cm}
    \begin{pmatrix} \|e_{ij}\|^2 & \hspace{-0.1cm}\langle e_{ij},e_{ik}\rangle  \\
     \langle e_{ij},e_{ik}\rangle & \hspace{-0.1cm} \|e_{ik}\|^2\end{pmatrix}\begin{pmatrix}  f(\mathbf{v}_j) - f(\mathbf{v}_i) \\ f(\mathbf{v}_k)- f(\mathbf{v}_i) \end{pmatrix}
\end{align*}
where the notation is as per the inset figure. 
 The divergence operator $\mathbf{D}\in\mathbb{R}^{|V| \times 3|F|}$ takes a 3D vector per triangle and returns a value per vertex. As the  adjoint of the gradient, it is discretized as : %
\begin{align}
\mathbf{D} = -\mathbf{A}^{-1}  \mathbf{G}^\top \mathbf{T}\,,
\end{align}
where $\mathbf{A}$ is a diagonal matrix of local area elements at each vertex (shaded area in the inset) and $\mathbf{T}$ is a diagonal matrix of triangle areas. 
These discretizations follow the standard formulations as described, e.g., in \cite{botsch2010polygon}.

To define these operators on point clouds, we base our approach on the construction of a graph $(\mathcal{V},\mathcal{E})$, built locally at each point in a $k$-neighborhood. The edges of the graph are weighted by a Gaussian kernel $w_{ij}=e^{-\frac{\|x_i-x_j\|_2^2}{\delta}}$\textcolor{red}{, for an appropriate $\delta$ depending on the point set density}. Functions $f: \mathcal{V} \to \mathbb{R}(\mathcal{E})$ are  defined on the nodes of the graph while vector fields $F: \mathcal{E} \to \mathbb{R}(\mathcal{V})$ are defined on the edges. The gradient and divergence operators are \sout{respectively} defined as: 
\begin{align}
    \mathbf{G}_{ij}&= w_{ij}( f_i - f_j) \ \ \ \forall (i,j) \in \mathcal{E}\\
    \mathbf{D}(F_i)&=\sum_{j \in (i,j) \in \mathcal{E}} \hspace{-0.2cm} w_{ij}(F_{ij}-F_{ji})
\end{align}

\sout{Finally,} In order to deal with low-quality meshes, we adopted the \sout{intrinsic triangulation} approach of \cite{sharp2019navigating}; a more detailed discussion with an experimental comparison is given in Appendix~\ref{sec:params}.

\subsection{Minimization of Eq.7}
The energy to minimize in line 3 of Alg.~\ref{alg:1} is convex, but non-differentiable. To efficiently solve it, 
 we discretize the problem and use the primal-dual hybrid gradient algorithm~\cite{chambolle:2011}, which is well-suited for {\textcolor{red}{problems of this form.}} %
The optimization procedure for an input signal $\mathbf{u}$ is summarized in Algorithm~\ref{PDHG}:
\begin{algorithm}
\SetAlgoLined
\SetKwInOut{Input}{input}\SetKwInOut{Output}{output}
\Input{$\mathbf{u}$, $\mathbf{q}$, $\sigma,\tau>0$, $\theta \in [0, 1]$}
\Output{minimizer $\mathbf{u}^\ast$ of Eq.~\eqref{eq:1}}
\Begin{
	\While{not converged}{
		$\mathbf{u}_{k+1}=prox_{\mathcal{G},\tau}(\mathbf{u}_k - \tau \mathbf{D} \mathbf{q}_k)$\\
		$\bar{\mathbf{u}}_{k+1}=\mathbf{u}_{k+1}+\theta(\mathbf{u}_{k+1}-\mathbf{u}_{k});$\\
		$\mathbf{q}_{k+1}=prox_{\mathcal{F}^*}(\mathbf{q}_k + \sigma \mathbf{G} \bar{\mathbf{u}}_{k+1})$
		}
\Return $\mathbf{u}^\ast= \mathbf{u}_{k+1}$
\caption{\label{PDHG}Primal-Dual Hybrid Gradient (PDHG)}}
\end{algorithm}
%
the proximal operator $prox_{\mathcal{F}^*}$ \sout{of $\mathcal{F}^*$} is the projection of each component of its input onto the unit $L^2$ ball: 
\begin{align}
    prox_{\mathcal{F}^*}(\mathbf{q}_{j,:})= 
    \begin{cases}
\mathbf{q}_{j,:} & \|\mathbf{q}_{j,:}\| \leq 1 \\
\frac{\mathbf{q}_{j,:}}{\|\mathbf{q}_{j,:}\|} & \text{otherwise}\end{cases}\,,
\end{align}
where each $\mathbf{q}_{j,:}\in\mathbb{R}^3$ is a vector at triangle $j$.
%
%
The proximal operator $prox_{\mathcal{G},\tau}$ \sout{with respect to the data fidelity term} has a closed-form solution given by:
\begin{align}
    prox_{\mathcal{G},\tau}(\mathbf{u})=\frac{\mathbf{u} + \frac{\tau}{\alpha}\mathbf{u}_0}{1+\frac{\tau}{\alpha}}\,.
\end{align}

We initialize the algorithm with $\mathbf{u}$ coming from the previous iteration in Algorithm~\ref{alg:1} and $\mathbf{q}=\mathbf{0}$, and stop the iterations when the absolute change in energy is below a small threshold. 
We set $\theta=0.5$, which appeared optimal in \sout{terms of} our numerical experiments. 

\section{Geometric TV flow}
\sout{In this Section,} We now study the evolution of the surface $\M$ \sout{itself} along the TV flow.
A direct way to do this is to evolve the $(x,y,z)$ vertex coordinates of $\M$; the same approach is also used to define, e.g., the mean curvature flow (MCF) and its variants \cite{Taubin:1995,Desbrun:1999,kazhdan2012can}. 
This approach, however, gives rise to two main issues.


First, diffusing the vertex coordinates naturally affects the metric of $\M$ at each diffusion step; this, in turn, modifies the functional $TV(f)=\int_\M \|\nabla_\M f\|$, since the integration domain $\M$ and the attached intrinsic operators now vary along the flow. 
 Recomputing the metric-dependent operators at each iteration of the algorithm is a possible solution, but can be highly inefficient.
 


\begin{figure}
\centering
      \includegraphics[width=\linewidth]{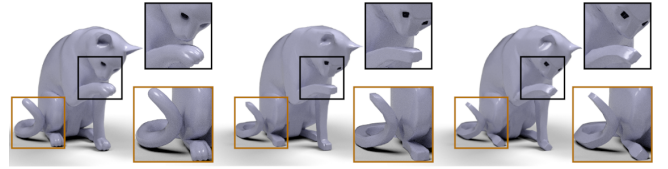}
  \vspace*{-0.7cm}
  \caption{\label{fig:cats}Diffusing vertex coordinates on the cat model. An anisotropic TV penalty will promote piecewise-flat solutions, where oscillations are minimized along the coordinate axes. Therefore, one gets different results by rotating the global reference frame; for the source model on the left, we show solutions \sout{of the flow} under two different axis rotations (middle and right respectively).}
\end{figure}

Secondly, one must take care of the choice of the TV regularization for $(x,y,z)$. In fact, for $f$ being the coordinate function on the manifold, $\nabla f(x)$ becomes a tensor for which a suitable norm $\|\nabla f(x)\|$ in the integral of the total variation needs to be defined. Depending on such a choice one can encourage different types of collaborative gradient sparsities \cite{Duran16}. For example, by choosing a fully anisotropic TV (i.e. summing the absolute value of all entries in $\nabla f(x)$), the transform will be orientation-dependent: the diffused vertex coordinates will tend to align to the global reference frame of $\mathbb{R}^3$, giving different results depending on the orientation of the initial shape; see Figure~\ref{fig:cats} for examples. While this latter point can be exploited for certain stylization tasks (as we show in Section~\ref{sec:apps}), it might be an undesirable effect in many other applications.

\begin{figure}
  \centering
\begin{overpic}[trim=0cm 0cm 0cm 0cm,clip,width=0.95\linewidth]{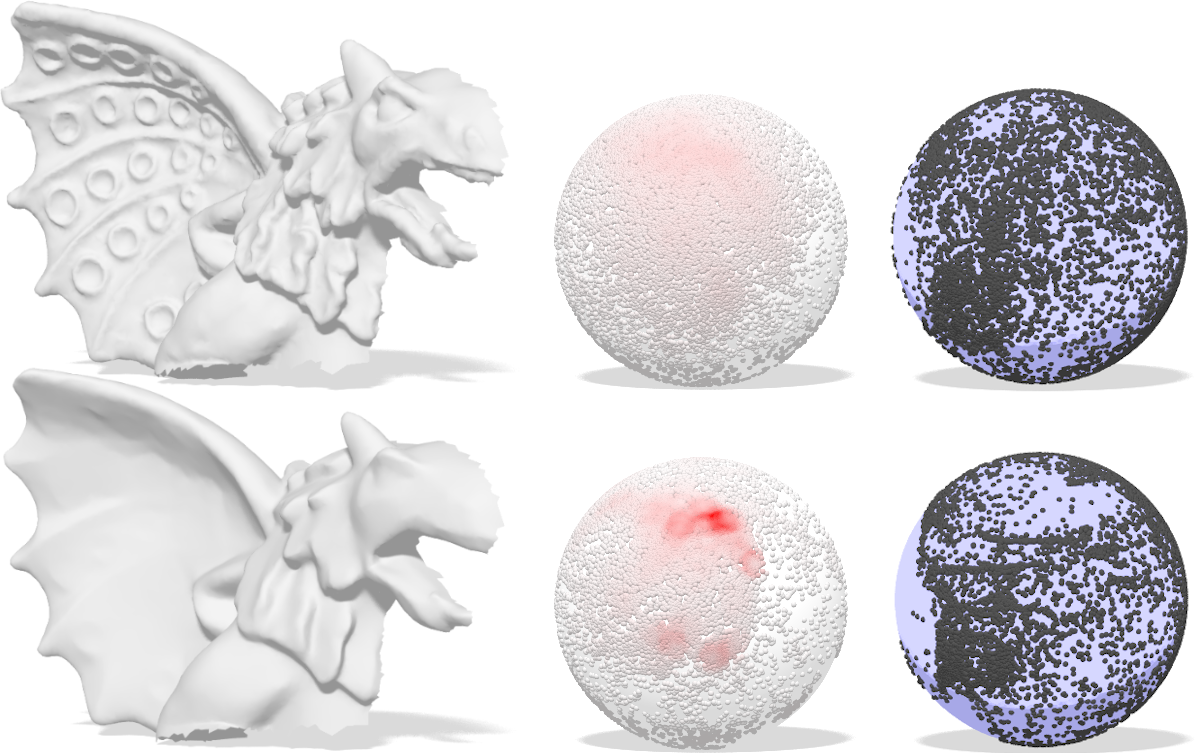}
\end{overpic}
  \caption{\label{fig:gargoyle}Our flow induces the formation of clusters in the normal field. The top row shows the original surface, the bottom row is the filtered result. {\em Middle}: the local density of the Gauss map (heatmap growing from white to red) illustrates the formation of clusters in normal space. {\em Right}: the coverage of $\mathcal{S}^2$ \textcolor{red}{(front and back sides visualized)} shows that the diffused normals are less spread across the sphere than in the original surface.\sout{; the other side of the sphere is visualized here for a more complete picture.}}
\end{figure}


\sout{Due to the aforementioned problem of changing meshes, we instead propose a different type of geometric flow as we will discuss in the next section.}

\subsection{Normal TV flow}

In this paper we advocate the use of normal fields for encoding the geometry. This choice has three main benefits: (1) the domain upon which the operators are defined remains fixed throughout the flow; (2) using normals endows us with {\em rotation invariance}; and (3) we get interesting connections with other existing flows.
Normal fields in connection with the TV functional have been considered before, e.g. in \cite{Zhang:2015,Zhong:2018}. However, these approaches do not consider or analyze the geometric flow as we do here, and do not provide a natural and interpretable multiscale decomposition.

\setlength{\columnsep}{7pt}
\setlength{\intextsep}{1pt}
\begin{wrapfigure}[5]{r}{0.40\linewidth}
\begin{center}
\begin{overpic}
[trim=0cm 0cm 0cm 0cm,clip,width=1.0\linewidth]{./images/buns.png}
\put(49.5,35){\footnotesize $\Gamma$}
\put(46,29){\footnotesize $\longrightarrow$}
\end{overpic}
\end{center}
\end{wrapfigure}
%
Our input signal is the normal vector field $\vec{n}_0$ over $\M$. The field $\vec{n}_0$ is identified by the image of the Gauss map $\Gamma : \M \rightarrow \mathcal{S}^2$, mapping points on the surface to points on the unit $2$-sphere (see inset).
The corresponding time continuous flow is:
\begin{align}\label{eq:npde}
    \begin{cases}
    \frac{\partial \Gamma}{\partial t}= ~\mathrm{div}_{S} \left(\frac{ d\Gamma }{\|d\Gamma\|}\right) &  \text{in  } \mathcal{S}^2,\\
    \Gamma(0,x) = \vec{n}_0 \,,
    \end{cases}
 \end{align}
 where $d \Gamma$ is the differential of the Gauss map \sout{(aka shape operator),} and $\mathrm{div}_S$ is defined with the metric of $\M$, but applied to tangent fields on $\mathcal{S}^2$.
By diffusing normals according to Eq.~\eqref{eq:npde}, \sout{the domain $\M$ stays fixed throughout the entire flow (thus, its metric does not need to be updated at each iteration), while} the point distribution over $\mathcal{S}^2$ changes to form well-separated clusters. Intuitively, the flow tends to align the directions of normal vectors to favor piecewise-constant solutions; see Figure~\ref{fig:gargoyle} for an example. \sout{of normal re-distribution induced by the flow;} To obtain the filtered embedding for $\M$, the diffused normal field is integrated as described further in this section.

\sout{
The clustering induced by the normal TV flow is {\em not} equivalent to running a clustering algorithm in normal space, since the flow is defined w.r.t. the metric on the {\em original} surface $\M$. A simple clustering of nearby normals would instead use the metric of $\mathcal{S}^2$.
}





Interpreting this approach from a differential geometric perspective provides us with a useful characterization of the normal TV flow in terms of surface curvature. In particular, the TV functional for the normal field $\vec{n}$ reads:
\begin{align}\label{eq:soscurv}
    TV(\vec{n})=
    \int_{\M} \| d \Gamma (x) \| dx=
    \int_\M \hspace{-0.05cm} \sqrt{k_1^2 + k_2^2} \, dx\,,
\end{align}
where  
 $k_1,k_2$ are the principal curvatures at each point.
 The last equality follows from the fact that $k_1,k_2$ are the eigenvalues of $d\Gamma$.
 
 
 
As we show below, the final expression in Eq.~\eqref{eq:soscurv} resembles the functional related to other known flows in the literature. We also see from the same expression that the TV energy of the normal field does {\em not} depend on the orientation of the normals in $\mathbb{R}^3$; therefore, the resulting flow is invariant to global rotations of $\M$. 


\vspace{1ex}\noindent\textbf{Implementation.}
By working with normal fields $\vec{n}$ we are moving to manifold-{\em valued} functions (the manifold being the unit sphere).
 In our tests we treat each $\vec{n}(x)$ as an element of $\mathbb{R}^3$, and add a normalization step projecting $\vec{n}(x)$ back onto $\mathcal{S}^2$ at each iteration.


In the discrete setting, the normal vector field $\mathbf{n}\in\mathbb{R}^{|F|\times 3}$ is constant within each triangle. 
 Its gradient (needed in Algorithm~\ref{PDHG}, line 5) is understood as the channel-wise jump discontinuity over each edge in each triangle, and the associated discrete operator can be assembled as a $|E| \times |F| $ matrix; see Appendix~\ref{sec:edgegrad} for details.
 
 \begin{figure}[t]
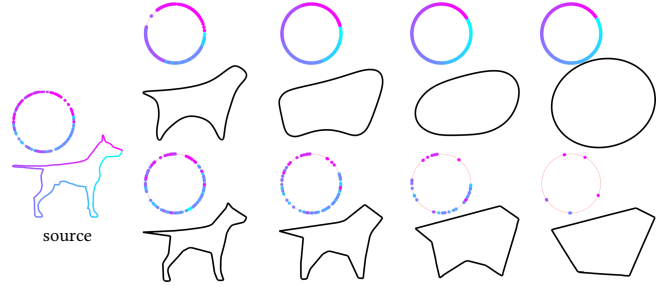

    \centering
\begin{overpic}[width=\linewidth]{images/dogs.png}
\put(4.7,6.4){\footnotesize source}
\end{overpic}
\caption{Comparison between MCF (top row) and normal TV flow (bottom row) on a 1D closed contour \sout{depicting} of a dog. The circles visualize the image of the Gauss map on $\mathcal{S}^1$ for each shape; the color code is defined as on the source shape (leftmost column). The MCF equally spreads the normals, eventually converging to a circle, while the TV flow forms well-separated clusters \sout{of normals (in this example,} (\sout{concentrated around} the 5 points on the circle), converging to a piecewise-flat shape.}
\label{fig:1dflow}
\end{figure}

\vspace{1ex}\noindent\textbf{Recovering vertex positions.}
To recover a new embedding from the filtered normals $\mathbf{n}$, we follow \sout{the approach of} \cite{Prada:2015}\sout{ and solve the optimization problem}:
\begin{align}\label{eq:poisson}
        \min_{\mathbf{v} \in \mathbb{R}^{|V|\times 3}} \| \mathbf{G} \mathbf{v} - \mathbf{w}\|_F^2 + \epsilon \|\mathbf{v}-\mathbf{v}_0\|_F^2\,,
\end{align}
where $\mathbf{v}_0$ are the old vertex coordinates, and $\mathbf{w}=\mathbf{Gv}_0- \langle \mathbf{Gv}_0,\mathbf{n} \rangle_r \odot \mathbf{n}$ is the orthogonal component of $\mathbf{G} \mathbf{v}_0$ with respect to $\mathbf{n}$ ($\langle\cdot,\cdot\rangle_r$ and $\odot$ operate row-wise).
%
%
Eq.~\eqref{eq:poisson} seeks the set of vertices whose gradient matches the vector field $\mathbf{w}$, while at the same time staying close to the initial vertices. \sout{The first term is inspired by the gradient-based energy of Yu:2004.} In our tests we set $\epsilon=10^{-4}$ to allow large deviations from the initial shape.
Solving Eq.~\eqref{eq:poisson} boils down to solving a screened Poisson equation, giving the sparse linear system:
\begin{align}\label{eq:screened}
        \mathbf{v}=(\epsilon \mathbf{A}- \mathbf{L})^{-1}( \epsilon \mathbf{A} \mathbf{v}_0 - \mathbf{D} \mathbf{w})\,.
\end{align}
Here, $\mathbf{L}=-\mathbf{DG}$ corresponds to the standard linear FEM discretization of the Laplace-Beltrami operator.

\subsection{Related curvature flows}
\sout{Here} We compare our \sout{normal TV} flow to other geometric flows used in graphics.

\vspace{1ex}\noindent\textbf{Mean curvature flow.}
The \textcolor{red}{normal} TV flow is closely related to the MCF, which describes the evolution of a surface under the minimization of the area functional (membrane energy):
\begin{align} \label{eq:areafnctl}
    E_A(u) = \int_\M dx \,,
\end{align}
%
where $u$ is now the $(x,y,z)$ embedding in $\mathbb{R}^3$ of the surface $\M$, and we assume $\M$ has no boundary. The relation to our flow emerges by approximating the membrane energy with the Dirichlet energy $E_A(u)\propto D(u)=\frac{1}{2}\int_\M \| \nabla u \|^2$, whose gradient flow is the PDE:
\begin{align}
    \frac{\partial u}{\partial t} = \Delta u \,.
\end{align}
Compared to Eq.~\eqref{eq:pde}, this PDE uses the linear operator $\Delta=\mathrm{div}\nabla$ instead of the nonlinear operator $\mathrm{div}\frac{\nabla}{\|\nabla\|}$. 

Denoting with $H = \frac{k_1 + k_2}{2}$ the mean curvature, and observing that $\Delta u = 2H \vec{n}$, the above PDE corresponds to 
  surface motion in the normal direction, with speed proportional to curvature $H$. Critical points of the functional are minimal surfaces, which for genus-0 \textcolor{red}{convex} shapes correspond to spheres.
  
\begin{figure}
\begin{overpic}[width=\linewidth]{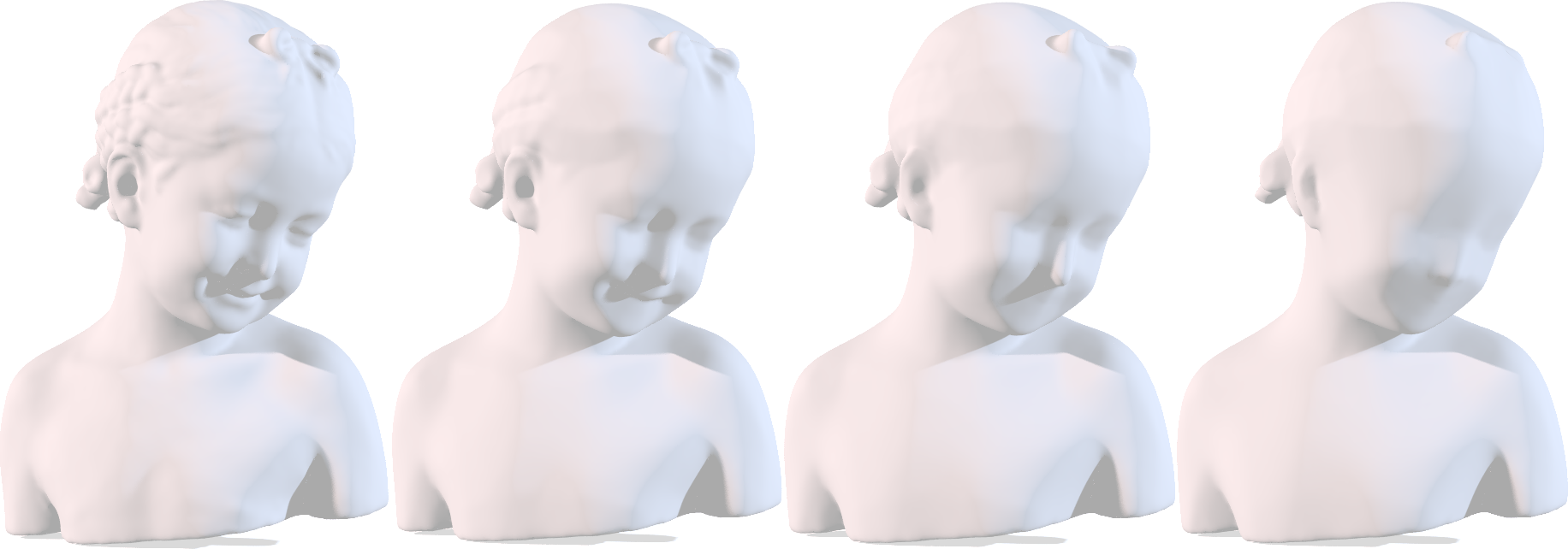}  
\end{overpic}
    \caption{\label{fig:bimba}Normal TV flow on the bimba\sout{ model}.  Sharp geometric features get smoothed in a ``vertical'' fashion, preserving the overall structure. For example, the chin and pony tail maintain their shape throughout, while the braid and nose slowly disappear without affecting the neighboring geometry.}
\end{figure}

In comparison, surface evolution along the normal TV flow collapses the surface to a flat shape, with the direction of motion perpendicular to the level sets of the normal vector field $\vec{n}$.
By looking at the effect of the flow on $\vec{n}$, the key difference is that the MCF tends to move points on $\mathcal{S}^2$ so as to keep them equally spaced, while the normal TV flow generates well-separated point clusters. We illustrate the comparison in 2D in Figure~\ref{fig:1dflow}; an example of our feature-preserving flow in 3D is given in Figure~\ref{fig:bimba}.

\vspace{1ex}\noindent\textbf{Willmore flow.}
Perhaps more interestingly, we observe that our functional $TV(\vec{n})$ resembles the Willmore (or thin plate) energy \cite{Bobenko:2005:}, \sout{which is} a higher-order version of the area functional, \sout{and is} defined as:
\begin{align}\label{eq:will}
    E_W(u) = \int_\M H^2 dx = \int_\M (k_1^2 + k_2^2) \, dx - \cancel{2\pi\chi(\M)} \,,
\end{align}
where again $u$ encodes the coordinate functions in $\mathbb{R}^3$. The last term is the constant Euler characteristic of $\M$, and thus it does not affect the flow.
Compared to Eq.~\eqref{eq:soscurv}, we see that the integrand only changes by a square root operation. In this sense, we interpret the normal TV flow as a sparsity-promoting variant (in the normal domain) of the Willmore flow. 

The linearization of Eq.~\eqref{eq:will} leads to a fourth-order bi-Laplacian diffusion equation:
\begin{align}
     \frac{\partial u}{\partial t} = \Delta^2 u \,.
\end{align}
Our problem also involves a fourth-order PDE with respect to the coordinates; however, it is hard to solve directly due to the nonlinearity of our operator. By phrasing the PDE in terms of the normals instead of the coordinates, we decouple the problem: first we integrate the TV flow in the space of normals, which corresponds to the {\em second}-order PDE of Eq.~\eqref{eq:npde}, and then we integrate again to recover the vertex coordinates via the screened Poisson equation. 

\begin{table}
  \begin{tabular}{lll}
    \toprule
    \textbf{Flow} & \textbf{Functional} & \textbf{PDE}\\
    \midrule
    MCF & $E_A(u) = \int dx $& $    \frac{\partial u}{\partial t} = \Delta u$ \\
    Willmore & $E_W(u) =\int (k_1^2 + k_2^2)$ & $   \frac{\partial u}{\partial t} = \Delta^2 u $
    \\
    TV & $E_{TV}(\vec{n}) = \int \hspace{-0.05cm} \sqrt{k_1^2 + k_2^2} \, dx$ & $\frac{\partial \Gamma}{\partial t}= ~\mathrm{div}_S \left(\frac{d\Gamma}{\|d\Gamma\|}\right)$\\
  \bottomrule
\end{tabular}
\caption{\label{tab:flows}Comparison between MCF, Willmore, and our normal TV flow in terms of functional involved and the associated PDE.}
\end{table}

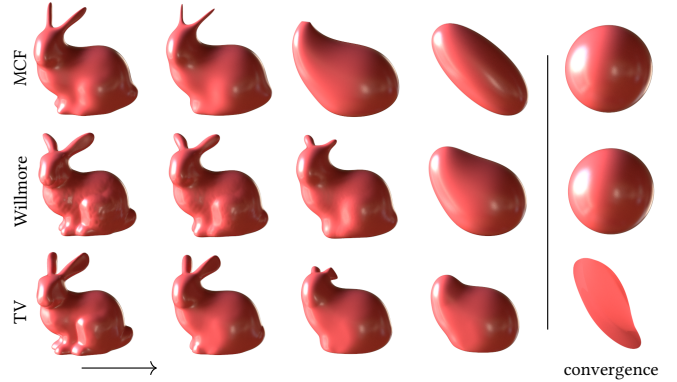
\begin{figure}
\begin{overpic}[width=\linewidth]{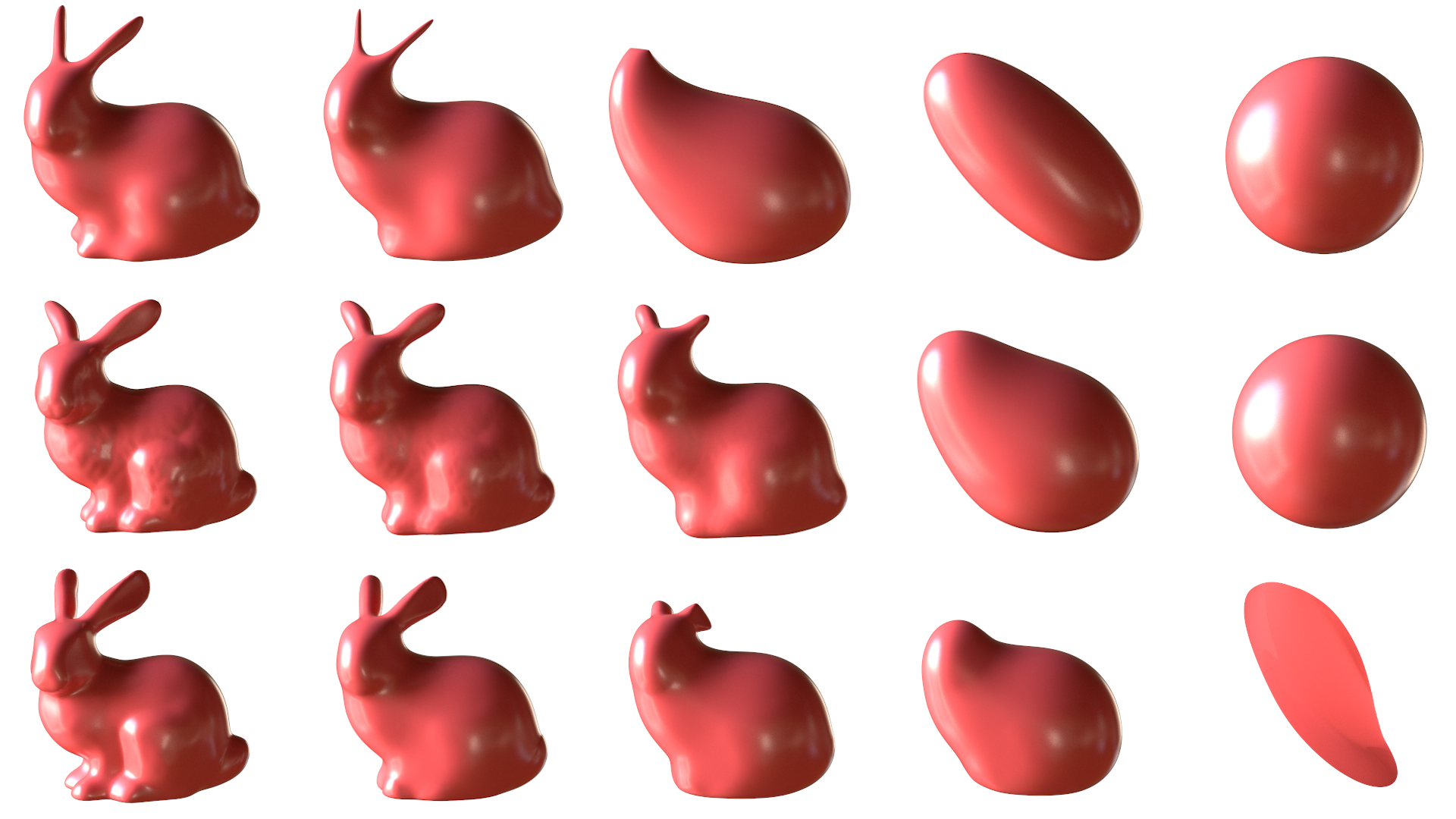}  
\put(-1.5,6){\rotatebox{90}{\footnotesize TV}}
\put(-1.5,21){\rotatebox{90}{\footnotesize Willmore}}
\put(-1.5,43){\rotatebox{90}{\footnotesize MCF}}
\put(8,-2){
\begin{tikzpicture}
\draw[->,line width=0.15mm](0,0) -- (1,0);
\end{tikzpicture}}
 \put(80.5,5){
\begin{tikzpicture}
\draw[-,line width=0.15mm](0,0) -- (0,3.65);
\end{tikzpicture}}
\put(84,-2){\footnotesize convergence}
\end{overpic}
    \caption{Evolution of the bunny surface along different geometric flows. Top to bottom: conformalized mean curvature flow \cite{kazhdan2012can}, conformal Willmore flow \cite{Crane:2013:RFC}, and our normal TV flow. The TV flow differs from the others as it preserves the underlying structure, and details are removed gradually. At convergence, we get a flat surface since the image of the Gauss map is clustered into a single point on $\mathcal{S}^2$.}
    \label{fig:3dflow}
\end{figure}

In Table~\ref{tab:flows} we summarize the three flows\sout{ to better highlight their analogies}. A qualitative comparison is shown in Figure~\ref{fig:3dflow}, \sout{for a fair comparison,} where we use stable variants of the existing flows that prevent \sout{numerical} singularities.
%


\subsection{The p-Laplacian operator}
\begin{figure}
    \centering
   \begin{overpic}[width=\linewidth,scale=0.25]{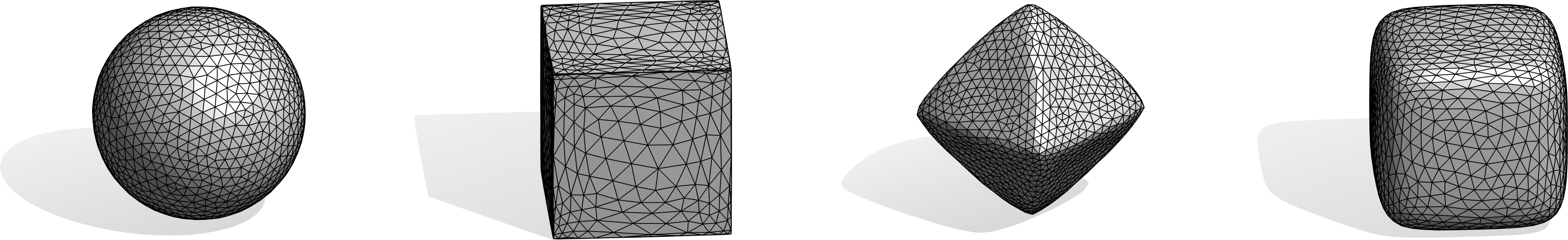}
    \put(9,17){\footnotesize $p=2$}  
    \put(37,17){\footnotesize$p=1$}  
    \put(62,17){\footnotesize$p=5$}  
    \put(89,17){\footnotesize$p=1.5$}
    \put(9.5,-3){\footnotesize MCF}
    \put(38.5,-3){\footnotesize TV}
    \end{overpic}
    \caption{Evolution of \textcolor{red}{the vertex coordinates} of the unit $L_2$ sphere along \sout{different} $p$-Laplacian flows, for different values of \sout{parameter} $p$. \sout{The input signal here are the vertex coordinates $(x,y,z)$;} The flows are simulated with the explicit Euler method. For $p=2$ the sphere remains a sphere, while it becomes a piecewise-flat surface with the TV flow ($p=1$).}
    \label{fig:plaplacian}
\end{figure}
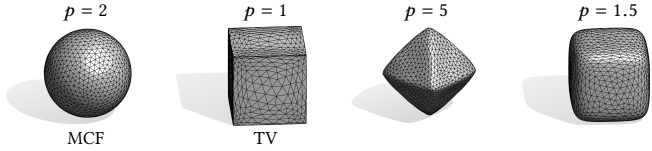

We remark here that both the Laplace-Beltrami operator $\Delta_\M$ and the \sout{nonlinear TV} subdifferential $\partial TV$ explored in this paper are special cases of a more general parametric operator, the $p$-Laplacian \cite{Lindqvist:2006}:
\begin{align}
    \Delta_p = \mathrm{div} (\|\nabla\|^{(p-2)} \nabla  )\,,
\end{align}
with the associated $p$-Dirichlet energy:
\begin{align}
    E_p(u) = \frac{1}{p}\int_\M \|\nabla u\|^p dx \,.
\end{align}
In particular, the $p$-Dirichlet energy with $p=1$ corresponds to the TV functional, while for $p=2$ we retrieve the standard Dirichlet energy.
By studying the solution for the evolution along the $p$-Laplacian flow of a unit ball, we observe convergence to a cube for $p=1$ (TV), to a sphere for $p=2$ (Dirichlet energy) and to an octahedron for $p \gg 2$, which is a consistent behavior with the definition of $p$-norms  (see Figure~\ref{fig:plaplacian}).
From the example we see clearly that the TV flow tends to piecewise-flat solutions, and further observe that other $p$-Laplacian flows carry other properties that may be worth investigating in the future. We refer to \cite{Buehler09} for a discussion on the eigenvectors of the $p$-Laplacian and to \cite{Cohen19} for an approach defining a $p$-Laplacian-based spectral decomposition in Euclidean spaces. 


\section{Applications}\label{sec:apps}
The TV spectral framework lends itself well to a variety of applications. In these, we follow the same computational procedure: 
\begin{enumerate}
\item Compute the forward TV transform of a given signal (possibly, the geometry itself);
\item Apply a filter to the resulting spectral components;
\item Compute the inverse TV transform.
\end{enumerate}

%


\noindent
Letting $u:\mathcal{M}\to\mathbb{R}^k$ be a $k$-channel signal on surface $\M$, $\phi_t$ be its spectral representation, and $I$ be a (possibly nonlinear) filtering operator, the filtered signal $\tilde{u}$ is computed as:
\begin{align}
   \tilde{u}= \int t I(\phi_t) dt \,.
\end{align}
Note that the above formula is meant to be generic, while in practice the type of TV regularization used to construct $\phi_t$ plays an important role as we shall see below.

We showcase our method on a range of relevant applications for graphics; for each application, we compare with other approaches from the state of the art to better position our method in terms of achievable quality. While these results are intended to demonstrate the flexibility of the framework, they are by no means exhaustive and could serve as a basis for follow-up work.

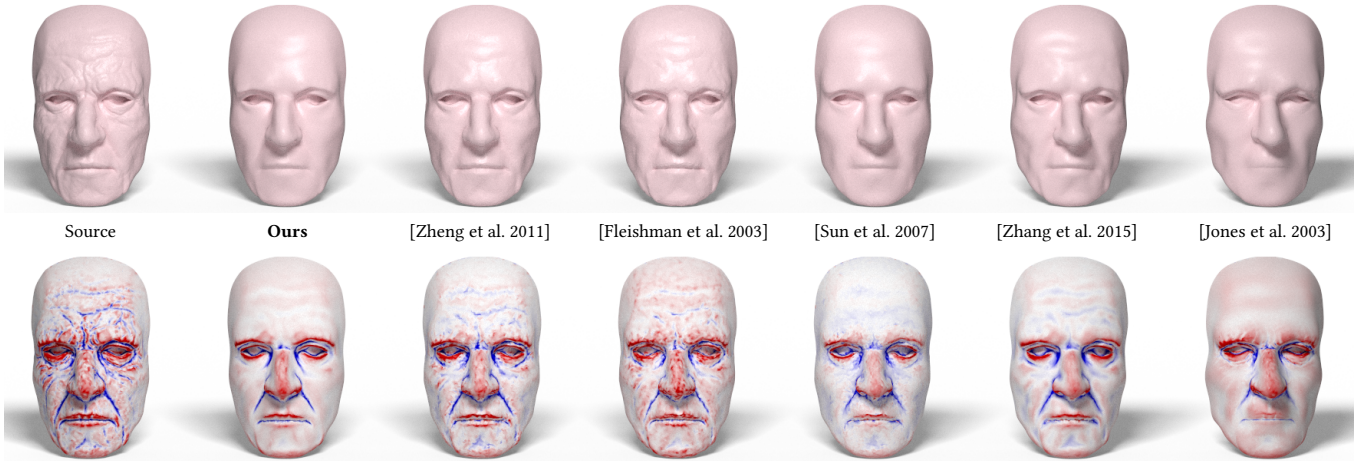
\begin{figure*}[t]
    \centering
    \begin{overpic}[width=\linewidth]{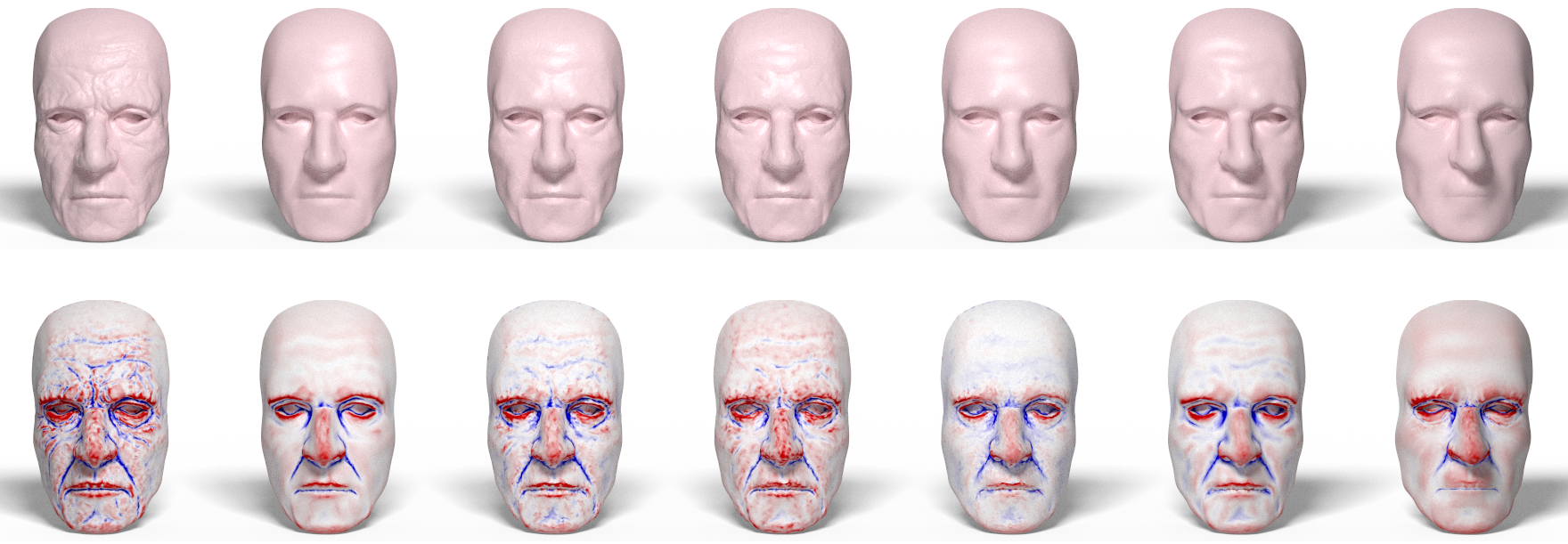}
    \put(4.5,17){\footnotesize{Source}}
    \put(19.5,17){\footnotesize{\textbf{Ours}}}
    \put(30,17){\footnotesize{\cite{Zheng:2011}}}
    \put(44,17){\footnotesize{\cite{fleishman03}}}
    \put(60,17){\footnotesize{\cite{Sun:2007}}}
    \put(73.5,17){\footnotesize{\cite{Zhang:2015b}}}
    \put(88.5,17){\footnotesize{\cite{Jones:2003}}}
    \end{overpic}
    \caption{\label{fig:removal}{\em Top row:} Example of high-frequency detail removal, implemented as a \sout{simple} low-pass filter on the spectral TV representation of the source shape (leftmost). We get a ``rejuvenation'' effect on the face model (second column), and compare our solution with state-of-the-art approaches, specifically tailored for edge-preserving mesh filtering (third to fifth column). {\em Bottom row:} To better highlight the differences, we plot the mean curvature (clamped to $[-30 , 30]$ and color-coded, growing from blue to red) on the filtered surfaces. Our method is the most effective at removing wrinkles and obtains sharper feature lines compared to other approaches, which either tend to over-smooth or retain spurious vertex noise. A quantitative comparison \sout{over this data} is reported in Table~\ref{tab:curvEnergy}.}
\end{figure*}

\begin{figure}
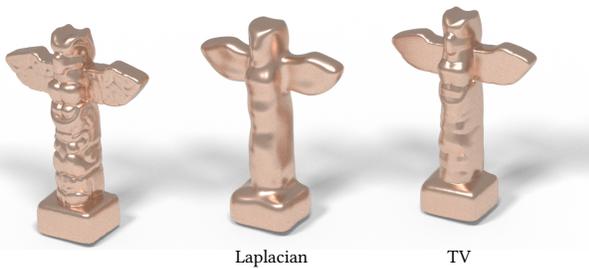


    \begin{overpic}[width=\linewidth]{./images/totem0.png}
    \put(44,-2){\footnotesize Laplacian}
    \put(77,-2){\footnotesize TV}
    \end{overpic}
  \caption{\label{fig:totem}Comparison between low-pass filtering of a source shape (left) \sout{in the Laplacian basis} using 300 Laplacian eigenfunctions (middle), and with our spectral TV representation using just \textbf{11 spectral components} out of 25 (right). The full shape has approximately $70k$ vertices. Laplacian-based smoothing correctly removes details, but also has a smoothing effect on the rest of the shape. In contrast, our approach correctly preserves the underlying structure.}
\end{figure}
\begin{table}
\resizebox{0.55\linewidth}{!}{
  \begin{tabular}{lc}
    \toprule
    \textbf{Method} & \textbf{TV Energy} \\
    \midrule
    Source & $43.412$ \\
     \textbf{Ours} & $\mathbf{26.55}$     \\
     \cite{Sun:2007} & $27.748$   \\
     \cite{Jones:2003} & $28.214$   \\
     \cite{Zhang:2015b} & $29.484$   \\
     \cite{fleishman03} & $31.918$   \\
     \cite{Zheng:2011} & $35.268$   \\
  \bottomrule
\end{tabular}
}
\caption{\label{tab:curvEnergy}Quantitative comparison between our TV spectral filtering method and the feature-preserving approaches from Figure~\ref{fig:removal}. For each method we report the TV energy of Eq.~\eqref{eq:soscurv}, which quantifies the amount of oscillations in the geometry after filtering; larger \textcolor{red}{values correspond to }\sout{numbers indicate the presence of} spurious noise.}
\end{table}

\subsection{Detail removal}
Perhaps the most straightforward application is the direct filtering of signals and geometric features living on the given surface. One such example is shown in Figure~\ref{fig:teaser}, where we apply a band-pass ``square window'' filter of the form:
\begin{align}
I_a^b(\phi_t) = \begin{cases}
\phi_t & t\in [a,b] \\
0 & \text{otherwise}\end{cases}
\end{align}
where $[a,b]$ is the range of time scales \sout{that are} left untouched by the filter; the normal field $\vec{n}$ is decomposed by applying Algorithm~\ref{PDHG} to  \sout{the isotropic total variation to} each component of $\vec{n}$ separately \textcolor{red}{and coupling the channels by projecting on the unit sphere at each step}. 
The above filter relies upon one main property of the TV decomposition: The separation of geometric details into different spectral bands, \sout{which is} ensured by the underlying assumption that most interesting shapes have a sparse gradient.
\sout{Despite its very basic form,} This simple filter \sout{can} already provides \sout{very} high-quality selective suppression of semantic detail while preserving sharp geometric features. In  Figure~\ref{fig:removal} we compare to five other top performing methods from the literature,\sout{which are} specifically tailored to \sout{tackle} edge-preserving filtering, while we show a comparison with Laplacian-based filtering in Figure~\ref{fig:totem}.  

\subsection{Detail transfer}
A remarkable application of the spectral TV framework is to transfer sharp geometric details between shapes. \sout{Let us be given} \textcolor{red}{Consider} two surfaces $\M,\mathcal{N}$ and a bijection $\pi:\mathcal{M}\to\mathcal{N}$ between them. Further, let $\phi_t$ and $\psi_t$ be the TV spectral representations of the normal fields on $\M$ and $\mathcal{N}$ respectively.
The idea is to transfer some components of $\phi_t$ to $\psi_t$ via $\pi$, and then compute the inverse transform of $\psi_t$ to obtain a new version of shape $\mathcal{N}$ with additional semantic details. 


For this to work, recall that $\phi_t : \M\to\mathbb{R}^3$ is a vector field on $\M$ for any given $t\in\mathbb{R}_+$, and similarly for $\psi_t:\mathcal{N}\to\mathbb{R}^3$.
Therefore, we can transfer details from $\M$ to $\mathcal{N}$ by evaluating the integral:
\begin{align}
\tilde{u}(x) &= \int t (  \psi_t(x) + I_a^b(\phi_t)\circ\pi^{-1}(x) ) dt \label{eq:tx1}\\
&= u(x) + \int_a^b t ( \phi_t \circ \pi^{-1}(x) ) dt \,,
\end{align}
where $u$ is the original normal field on $\mathcal{N}$, $\tilde{u}$ is the newly synthesized normal field, and $I_a^b$ is a band-pass filter that selects the desired features to borrow from $\M$. The integral in Eq.~\eqref{eq:tx1} is a direct application of the reconstruction formula of Eq.~\eqref{eq:inv} to an additively updated spectral representation $\psi_t$. The key feature of this procedure is that, despite its simplicity, it enables a form of {\em selective} transfer of details. Importantly, the details must not be necessarily localized on the surface, as long as they are well localized in scale space.

\begin{figure}
    \centering
\begin{overpic}[width=0.9\linewidth]{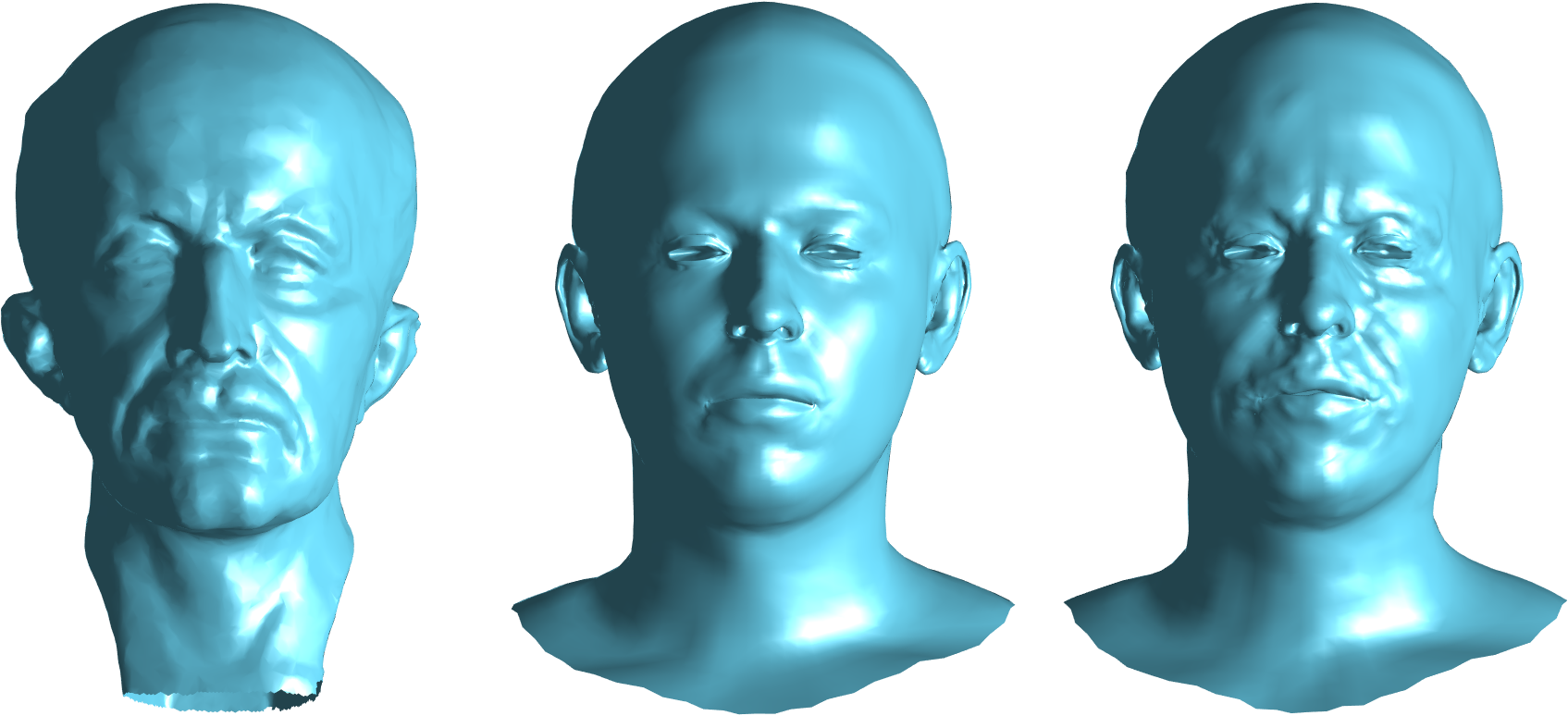}
\put(10,-3.5){\footnotesize{Source}}
\put(45,-3.5){\footnotesize{Target}}
\put(80.5,-3.5){\footnotesize{Result}}
\end{overpic}
    \caption{\label{fig:transfer}Example of detail transfer. The wrinkles on the source shape $\mathcal{X}$ (left) are captured with a high-pass filter in the spectral TV decomposition of $\mathcal{X}$, added to the spectral representation of the target (middle), and finally resynthesized (right). \sout{The clean separation between geometric features that is provided by} The TV decomposition allows to inject tiny {\em geometric} details into the target, without affecting the other bands (e.g., nose, mouth and ears are not affected by the transfer). Note that source and target shape do {\em not} have the same mesh connectivity, nor the same number of vertices.}
\end{figure}

In Figure~\ref{fig:transfer} we show an example of detail transfer.  In Figure~\ref{fig:coating} we compare our approach to the Laplacian-based detail transfer approach of \cite{Sorkine:2004}, where differences between Laplacian coordinates (i.e. differences between normals scaled by mean curvature) are transferred between shapes. For these tests, the map $\pi$ was computed using the method of \cite{zoomout} starting from $15$ hand-picked landmark correspondences.

\subsection{Stylization}\label{sec:Stylization}
\begin{figure}
    \centering
    \begin{overpic}[width=0.85\linewidth]{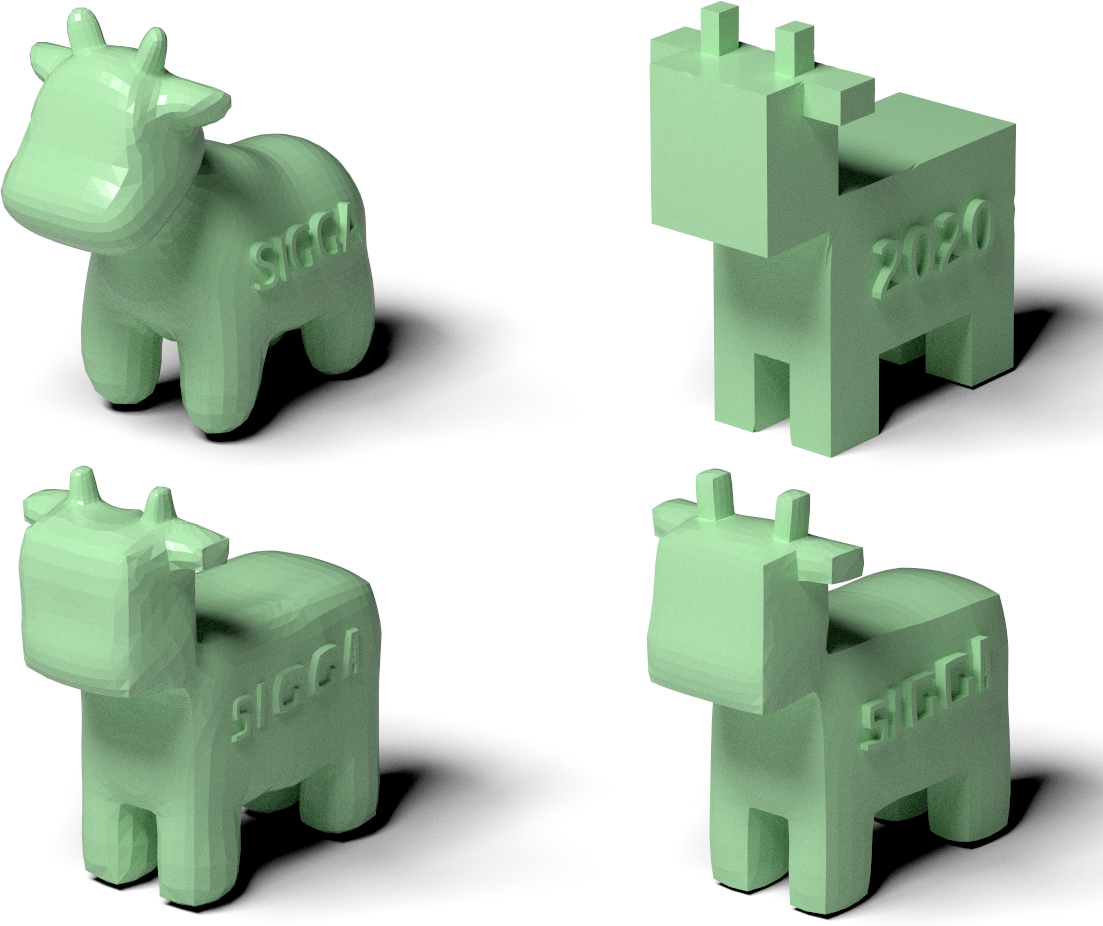}
\put(17,84){\footnotesize{Source}}
\put(9,-2.3){\footnotesize{\cite{Sorkine:2004}}}
\put(74,-2.3){\footnotesize \textbf{Ours}}
\put(73,84){\footnotesize{Target}}
\end{overpic}
    \caption{\label{fig:coating}{\em Replacing} the inscription from a source (top left) to a target model (top right), using Laplacian coating transfer \cite{Sorkine:2004} (bottom left), and our detail transfer approach (bottom right). The smoothing applied to the target causes the former approach to fail at preserving edges, while the transfer of high frequencies in the TV spectrum preserves them. The target model was obtained by applying our cubic stylization method of Section \ref{sec:Stylization} to the source model before superimposing the inscriptions.}
\end{figure}
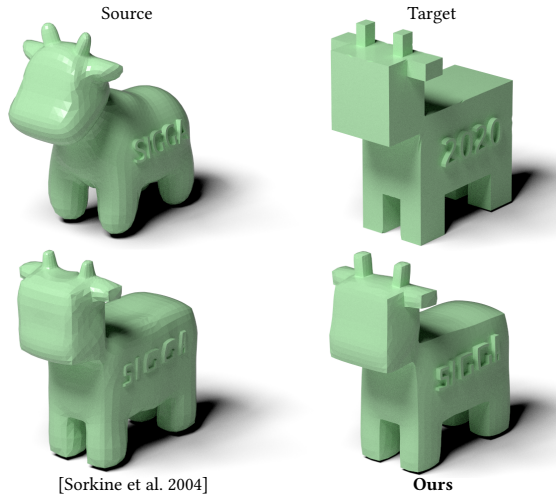

\begin{figure*}
    \centering
    \begin{overpic}[trim=0cm 0cm 0cm 0cm,clip,width=0.95\linewidth]{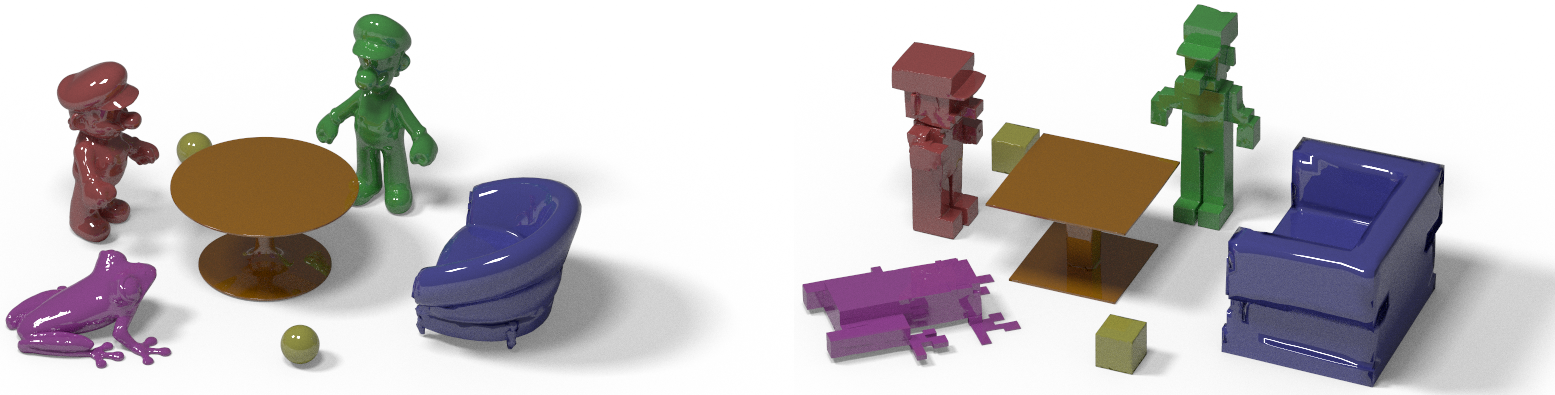}
    \end{overpic}
    \caption{\label{fig:cubic}Example of cubic stylization. The scene on the left is evolved along the {\em anisotropic} normal TV flow (i.e., each dimension of the normal vectors is treated as a separate scalar function, and these are regularized jointly via the sum of anisotropic TV penalties), which forces the normals to align with the global coordinate axes; the final result is shown on the right. \sout{The scene composition is for visualization purposes only;} Each shape was processed individually, and then placed in the scene.}
\end{figure*}

We further exploit some basic properties of the TV flow for artistic rendition of 3D shapes. In particular, we show how with a proper choice of the \textcolor{red}{TV functional} \sout{total variation of $u\in BV(\M)$}, we get the desired effect of mimicking the stylistic features typical of voxel art. A similar effect of ``cubic stylization'' was achieved in \cite{huang2014} \sout{with the different intent of} for constructing polycube maps for texturing, and in \cite{cubic} for modelling purposes. These approaches minimize an as-rigid-as-possible deformation energy with $L_1$ regularization on the normal field. \sout{While our approach is different, it shares some generic properties with these techniques; namely, the modified shapes do not need any remeshing, hence they} \textcolor{red}{Similarly to these methods,} our modified meshes retain any attribute defined on the original shape (e.g. UV coordinates for textures). Further, since in our case the stylization is the result of an evolution process, this can be stopped at any time to attain the desired sharpness level.
A simple \sout{and immediate} way to get cubic stylization \textcolor{red}{is to consider} \sout{consists in considering} the $(x,y,z)$ coordinates as separate scalar functions regularized via the sum of anisotropic TV penalties. 
This causes the surface to flatten during the diffusion, and eventually shrink to a single point at convergence. Due to the strict dependence on the global reference frame, different orientations of the surface in $\mathbb{R}^3$ will give different results, as in \sout{the approaches of} \cite{huang2014,cubic}. 
\textcolor{red}{Shrinkage can be avoided by diffusing the normal vector field instead of the vertex positions.} \sout{A different perspective is provided again by considering the normal vector field as our input signal, this way constraining the normals (instead of the vertices) to align to the global coordinate axes.} 
 \sout{The main benefit of working with normals is that there will be no flattening or shrinkage of the reconstructed surface, since} \textcolor{red}{This way,} the flow only modifies the normal directions, and solving the screened Poisson equation~\eqref{eq:screened} approximately preserves the volume.
 See Figures~\ref{fig:coating} and~\ref{fig:cubic} for some of our stylization examples.


\subsection{Denoising and enhancement}
%
Finally, we demonstrate the application of the spectral TV framework to more classical tasks of geometry processing, namely surface denoising and enhancement. For these tests, we use both triangle meshes and point clouds.
\textcolor{red}{Once the spectral decomposition is computed, filtering and reconstruction can be performed in real time, as shown in the accompanying video.}

\sout{For these applications (as well as for detail removal above), once a spectral decomposition is performed, the filtering and reconstruction operations are especially efficient and can be carried out at interactive times. This is particularly useful in view of interactive applications, as shown in the accompanying video.}

\begin{figure}
    \centering
    \begin{overpic}[width=0.85\linewidth]{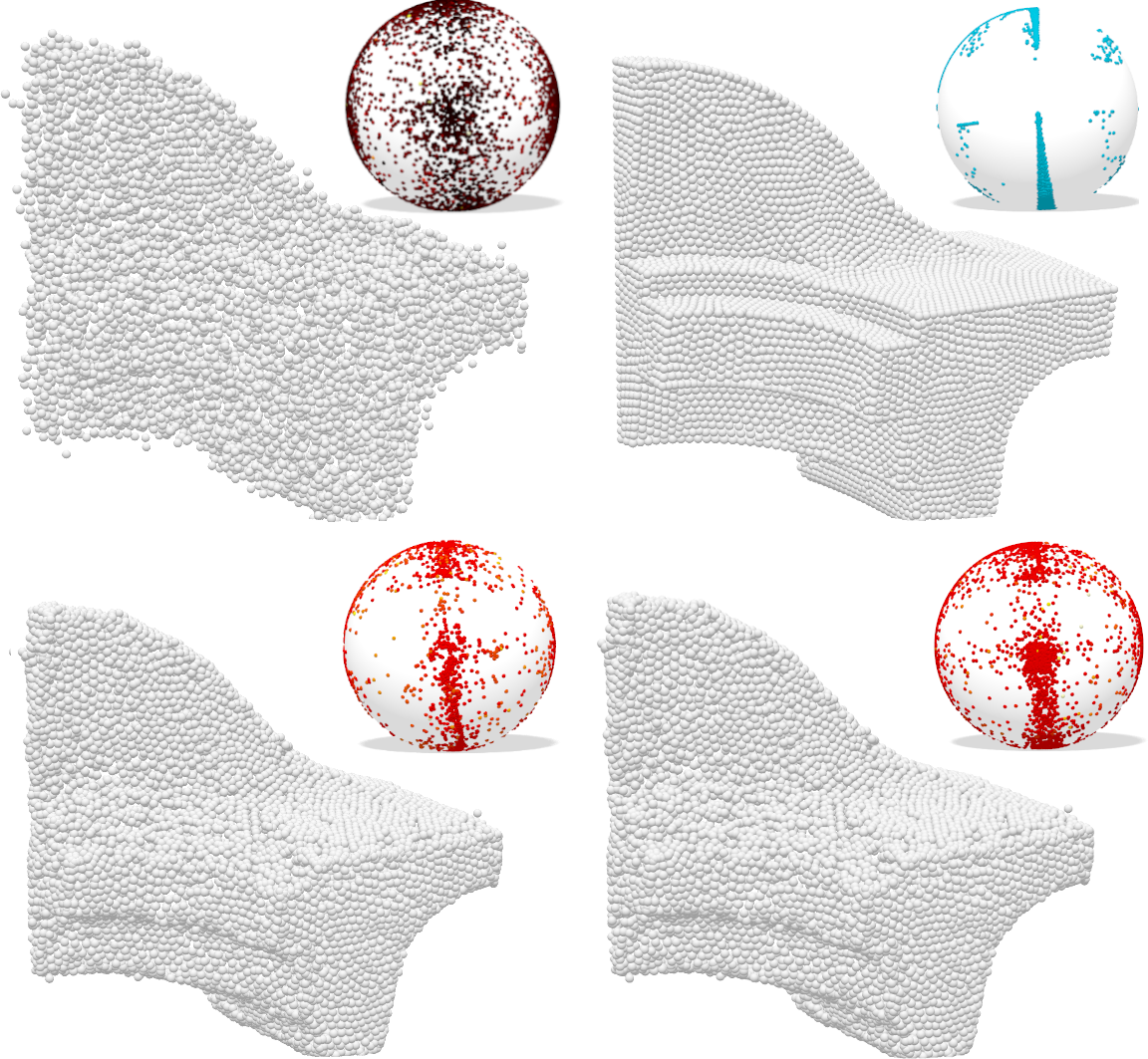}
\put(14,90.5){\footnotesize{input 0.059}}
\put(14.5,-2.3){\footnotesize \textbf{Ours 0.022}}
\put(50,-2.3){\footnotesize{\cite{Digne:2017} 0.027}}
\put(67,90.5){\footnotesize{ground truth}}
\end{overpic}
    \caption{\label{fig:denois}The oriented point cloud on the top left (obtained by additive Gaussian noise on the vertex coordinates\sout{ of the ground truth}) is denoised by \sout{computing the spectral TV representation of its normal field, and} applying a low-pass filter \textcolor{red}{to the spectral TV representation of its normal field} (bottom left). On the bottom right we compare to Digne \sout{the recent bilateral filtering approach for point clouds of }\shortcite{Digne:2017}. At the top-right of each point cloud, we plot its normal field on the unit sphere. Points are color-coded according to their cosine similarity \textcolor{red}{(light to dark red, with values reported in the figure)} with the ground truth normals \sout{(small error is light red, large error is dark red)}\sout{ the ground truth normals are colored with} (in solid cyan)}
\end{figure}

\vspace{1ex}\noindent\textbf{Fairing.} \sout{Especially} When dealing with point clouds, denoising and fairing tasks are notoriously challenging due to the difficulty of distinguishing feature points from outlier noise. This confusion often requires pre-processing steps to localize sharp features and filter out the rest. In our setting, the edge-aware property of the TV functional allows us to automatically damp oscillations due to noise, while at the same time retaining the true details of the underlying surface.
We formulate denoising \sout{quite naturally,} as a simple low-pass filter in the spectral TV domain. The input signal is the normal field $\vec{n}:\M\to\mathbb{R}^3$; when $\M$ is discretized as a point cloud, normals can be estimated via total least squares \cite{mitra2003estimating}. See Figure~\ref{fig:denois} for an example of denoising of point clouds, where we compare with the recent bilateral filtering approach of \cite{Digne:2017}.

\begin{figure}
    \centering
\begin{overpic}[width=0.95\linewidth]{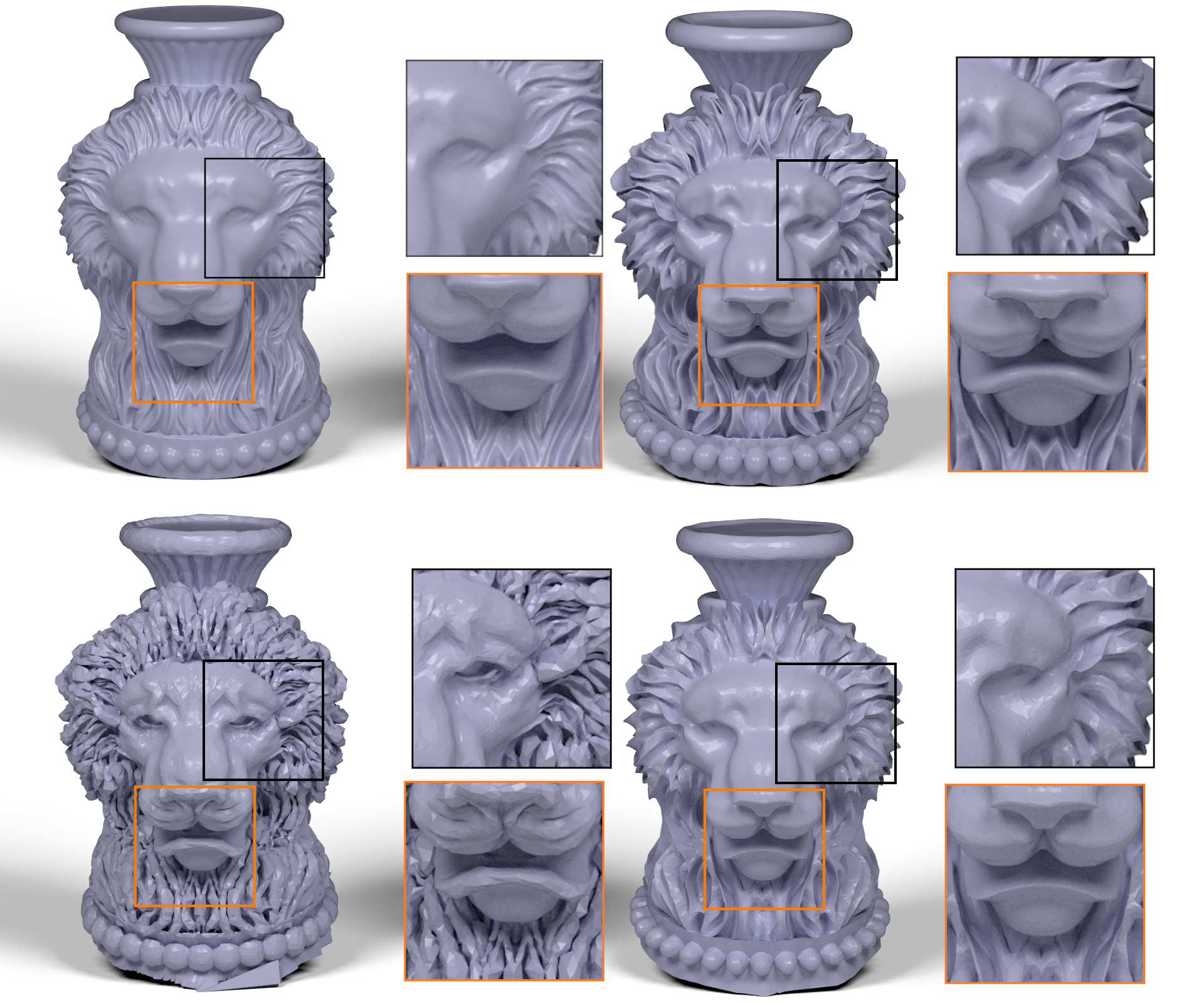}
\put(23,84.5){\footnotesize{Source}}
\put(72,84.5){\footnotesize \textbf{Ours}}
\put(16,-2.5){\footnotesize{\cite{Chuang:2016:GPO}}}
\put(54,-2.5){\footnotesize{\cite{Chuang:2016:GPO}} w/ smoothing}
\end{overpic}
    \caption{\label{fig:lion}{\em Top row:} Spectral TV enhancement of a lion-shaped vase. Assigning a larger weight to the high-frequency spectral components has an effect of pronouncing the details of the 3D model, while preserving the sharp features. {\em Bottom row:} Comparison with \cite{Chuang:2016:GPO}. Their method comes in two variants. On the left, their direct sharpening results in a noisy mesh. On the right, feature enhancement is alternated with smoothing 3 times, at the cost of detail loss (e.g. around the whiskers).}

\end{figure}
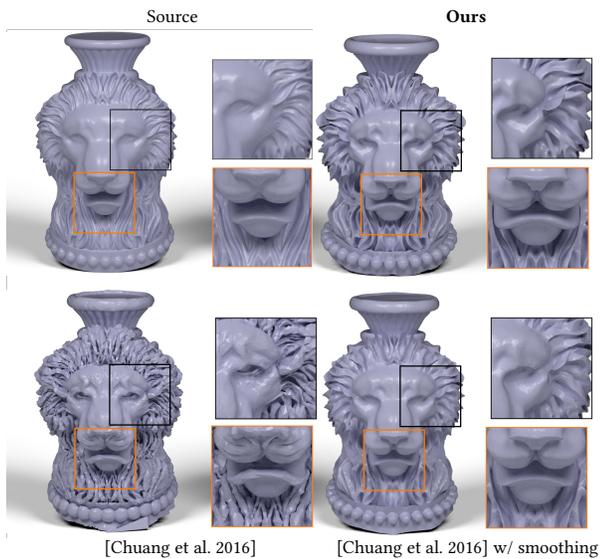

\vspace{1ex}\noindent\textbf{Enhancement.} While denoising can be seen as a suppression of high-frequency components in the spectral representation of $\vec{n}$, one can similarly obtain feature enhancement by {\em amplifying} those components; see Figure~\ref{fig:lion} for an example. Here we compare to the \sout{recent} gradient-based mesh processing approach of \cite{Chuang:2016:GPO}, which implements feature sharpening and smoothing \sout{(which are alternated in their pipeline) }by scaling the gradient of coordinate functions and solving for new coordinates via a Poisson equation. 
Conceptually, \sout{the main difference with our approach is in that} our spectral TV  representation allows more accurate control on specific features, since these are well separated in the spectrum; differently, the Laplacian-based approach adopted in \cite{Chuang:2016:GPO} encodes geometric features by spreading them out across the entire spectrum.

\vspace{1ex}\noindent\textbf{Localized operations.} We conclude by mentioning that our framework also supports localized operations by means of masks defined over regions of interest, which could be useful in view of an interactive application for mesh editing (see the supplementary video). Let $R\subset\M$ be a region of $\M$, and let $I$ be an arbitrary spectral filter. Then, for a signal $u$ with spectral components $\phi_t$, localized spectral filtering can be carried out simply by evaluating:
\begin{align}
    \tilde{u}(x) = \rho(x) \int_\M t I(\phi_t(x))  dt\, + (1-\rho(x)) u(x),
\end{align}
where $\rho:\M\to [0,1]$ is an indicator function for $R$. 




\section{Discussion and conclusions}
\sout{In this paper} We presented a new spectral processing framework for surface signals and 3D geometry\sout{(discretized as either triangle meshes or point clouds)}. The framework is based on ideas developed in the last decade within the nonlinear image processing community, but whose extension and application to geometric data \sout{and graphics tasks} have \sout{largely} been lagging behind to date. The main contribution of this paper is to show how these ideas can serve as valuable tools for computer graphics, and to relate the resulting observations and algorithms to more well known paradigms such as Laplacian-based spectral approaches and shape fairing via geometric flows.
Probably the most exciting aspect of the framework lies in its interpretability in terms of spectral components with an intuitive canonical ordering. Together with the \sout{discontinuity}\textcolor{red}{jump}-preserving property, \sout{the net result is} \textcolor{red}{one gets} the formation of frequency bands that tend to cluster together features with similar semantics. For example, the \sout{intertwining hair pattern} \textcolor{red}{hair strands} on a head model will be captured in a higher-frequency band than the ears; both will be captured sharply.
The general approach to mesh filtering that we described also retraces the steps of classical pipelines adopted in the linear case (transform -- filter -- resynthesize), making our nonlinear framework more accessible and of potentially broader impact.


\subsection{Limitations and future work}\label{sec:limitations}
\sout{Our method does have its limitations.} From a practical perspective, perhaps our weakest point is the need to define a schedule \sout{according to which} for the \sout{regularization} parameter $\alpha$ \sout{during the decomposition}, which determines the resolution of the spectral representation, \sout{in turn} influencing the design of filters. While in most cases an exact tuning \sout{of this parameter} is not required, it might be possible to define an adaptive scheme to automatically tune the scale of $\alpha$ to the feature landscape of the signal. Similarly, \sout{apart from the straightforward application of band-pass filters as we did in our tests,} a more sophisticated filter design to achieve complex effects is possible \sout{, but far from trivial} with \sout{To this end,} the adoption of learning-based techniques to predict task-specific filters \sout{may be highly beneficial}, as done \sout{for instance} in \cite{moeller2015learning} for image denoising. More generally, the integration \sout{this framework} within geometric deep learning pipelines is a promising direction that we are eager to pursue.
We also anticipate that the study of the more general $p$-Laplacian operator on surfaces \sout{will likely} \textcolor{red}{might} lead to a richer class of algorithms for geometry processing. One promising idea is to define the operator adaptively, \sout{by} selecting a value for $p$ depending, \sout{on certain geometric properties of the shape} e.g., on curvature. Finally, a more in-depth analysis of the normal TV flow as a robust alternative for mesh fairing is essential, due to its \sout{remarkable} feature preservation properties and its \sout{strict} connection to other \sout{common} geometric flows.



\begin{acks}
The authors gratefully acknowledge the anonymous reviewers for the thoughtful remarks, and Matteo Ottoni, Giacomo Nazzaro and Fabio Pellacini for the artistic and technical support. MF and ER are supported by the ERC Starting Grant No. 802554 (SPECGEO) and the MIUR under grant ``Dipartimenti di eccellenza 2018-2022'' of the Department of Computer Science of Sapienza University.
\end{acks}

%
%
%
\bibliographystyle{ACM-Reference-Format}
\bibliography{main}
%
\appendix
\section{Mathematical derivations}

\subsection{Proof of Theorem \ref{th:perimeter}}
\label{apx:tper}

For functions defined on Euclidean spaces (as opposed to surfaces, which we prove here), we refer to \cite{Fleming:1960}.

\begin{thm2}
Let E $\subset \M$ be a measurable closed subset of surface $\M$, with smooth boundary $\partial E$, and let $\chi_E$ be its indicator function. Then:
\begin{align*}
    TV(\chi_E)= \mathrm{length}(\partial E)\,.
\end{align*}
\end{thm2}

\begin{proof}
We split the proof by first proving $TV(\chi_E) \leq \mathrm{length}(\partial E)$ and then $TV(\chi_E) \geq \mathrm{length}(\partial E)$.
Let $V$ be the tangent vector field that attains the supremum in the weak definition of TV (see Eq.~\eqref{dfn:dualTV}), and let $\vec{n}$ be the unit normal vector to $\partial E$. We have: 
\begin{align*}
TV(\chi_E)=  \int_\M \chi_E \mathrm{div} V dx =|\int_\M \chi_E \mathrm{div} V dx| =|\int_E \mathrm{div} V dx|
\end{align*}
We now apply the divergence theorem and Cauchy-Schwarz inequality:
\begin{align*}
& | \int_{\partial E} \langle V  ,  \vec{n} \rangle d\ell ~| \leq \int_{\partial E} |\langle V , \vec{n} \rangle| d\ell  \leq\int_{\partial E} 1 ~ d\ell = \mathrm{length}(\partial E) \,.
\end{align*}
For the other direction, we pick an arbitrary vector field $U$, s.t. $|U| \leq 1$ and $U=\vec{n}$ on $\partial E$. This vector field is guaranteed to exist since $\partial E$ is smooth. Then, applying divergence theorem:
\begin{align*}
   \int_\M \chi_E \mathrm{div} U dx =\int_{E}  \mathrm{div} U dx=\int_{\partial E} \langle U  , \vec{n} \rangle d\ell= \int_{\partial E} 1 ~d\ell= \mathrm{length}(\partial E)\,.
\end{align*}
Finally, from Eq.~\eqref{dfn:dualTV} we get that:
\begin{align*}
   \mathrm{length}(\partial E)= \hspace{-0.1cm} \int_\M \hspace{-0.1cm} \chi_E \mathrm{div} U dx \leq  \hspace{-0.1cm} \sup_{\substack{V :\M\to\mathcal{T}\M\\ |V|_\infty\leq 1}} \int_\M \hspace{-0.1cm} \chi_E \mathrm{div} V dx = TV(\chi_E)\,.
\end{align*}
Therefore, it follows that: 
\begin{align*}
    TV(\chi_E)= \mathrm{length}(\partial E)\,.
\end{align*}
\end{proof}

\section{Algorithmics}

\subsection{Inverse scale space}\label{apx:inverse}
The inverse scale space method starts from the projection of the input signal $u_0$ into the kernel of the TV (yielding the mean of $u_0$) and converges in an inverse fashion to $u_0$ \sout{. This is governed by} \textcolor{red}{according to} the PDE: 
\begin{align}\label{eq:iss}
    \begin{cases} \partial_t p(t) = u_0-u(t) \\
    p(t) \in \partial TV(u(t)) \\
    p(0)  = 0 \end{cases}
\end{align}
The inverse flow shares many properties with the forward TV flow and has finite time extinction (\sout{at least} in finite dimensions after inverting time $t\rightarrow 1/t$ to account for the inverse nature of the flow, see \cite[Proposition 5]{Burger:2016}). 
The full algorithm is summarized below.
Other than reaching stationarity in an inverse fashion, the main difference with the forward flow is that the solution has a piecewise-constant behavior in time, as opposed to the linear behavior for the forward case, as seen in Section~\ref{section:4:3}.
This implies that it is sufficient to take a single derivative of the primal variable $u(t)$ \sout{in order} to compute a spectral component $\phi_t$ (line 7 of Algorithm~\ref{alg:inv}). 
\sout{For the Euclidean setting, we refer to \cite{Burger:2016}. for more details on spectral decompositions using the inverse flow.}
\begin{algorithm}
\SetAlgoLined
\SetKwInOut{Input}{input}\SetKwInOut{Output}{output}
\Input{signal $u_0$, max diffusion time $\alpha$, no. of time steps $N$}
\Output{spectral representation $\phi_t$}
\Begin{
    $v(0) \xleftarrow{} 0$ \\
    $u(0) \xleftarrow{}  \mathrm{mean}(u_0)$\\
	\For{$t=0:N$}{
		$u(t+1)  \xleftarrow{Alg. \ref{PDHG}} \arg\min_u TV(u) + \frac{1}{2 \alpha} \| u- (u_0 + v(t))\|_2^2 $\\
		$v(t+1)  \xleftarrow{}  v(t) - (u_0 - u(t+1))$ \\ 
		output $\phi_t(t) \xleftarrow{} u(t+1)-u(t)$ 
			}
\caption{\label{alg:inv}Inverse scale decomposition}}
\end{algorithm}

\subsection{Forward vs inverse flow and the role of $\alpha$}\label{apx:inversecomp}
\sout{In this section} We show the discretization of the forward and inverse flows and the computation of the related time scales, to provide a  better understanding of the inverse relation w.r.t. time between the two and to give better insight on the role of the regularization parameter $\alpha$. \sout{Full derivations are provided since, despite (and perhaps due to) their simplicity, these do not appear in the literature to our knowledge.}

\vspace{1ex}\noindent\textbf{The inverse flow}
is given by  Eq.~\eqref{eq:iss}. 
 Its \sout{implicit Euler} discretization \sout{of this equation} yields:
\begin{align*}
 &\frac{p(t^{k+1})-p(t^k)}{\Delta t^{k}} = u_0-u(t^{k+1})\\
  \Leftrightarrow \qquad & 0 = \Delta t^{k}\left(u(t^{k+1})-u_0 - \frac{1}{\Delta t^{k}}p(t^k)\right) +  p(t^{k+1}) 
\end{align*}
which -- due to $ p(t) \in \partial TV(u(t))$ -- is the optimality condition to:
$$ \min_u \frac{\Delta t^{k}}{2}\left\|u- u_0 - \frac{1}{\Delta t^{k}}p(t^k) \right\|_2^2 +  \text{TV}(u)\,. $$
On the other hand, the update formula in our Algorithm~\ref{alg:inv} solves:
$$ u_{new}  \xleftarrow{PDHG} \arg\min_u \frac{1}{2 \alpha} \| u- (u_0 + v(t))\|_2^2 +\text{TV}(u)\,. $$
Therefore, we identify $\alpha$ with $1/\Delta t^{k}$, and $v(t)$ with $\frac{1}{\Delta t^{k}}p(t^k)$. For the latter it is important to look at the update equation for $p(t^{k+1})$: \sout{coming from Eq.~\eqref{eq:iss}:}
$$  p(t^{k+1})  = p(t^k) - \Delta t^{k} (u_0-u(t^{k+1})) .$$ 
Dividing this equation by $\Delta t^{k+1}$ yields:
\begin{align*}
\frac{1}{\Delta t^{k+1}}  p(t^{k+1})  &=\frac{1}{\Delta t^{k+1}} p(t^k) - \frac{\Delta t^{k}}{\Delta t^{k+1}} (u_0-u(t^{k+1}))\,, \\
&=\frac{\Delta t^k}{\Delta t^{k+1}} \left(\frac{1}{\Delta t^{k}} p(t^k)\right) - \frac{\Delta t^{k}}{\Delta t^{k+1}} (u_0-u(t^{k+1}))\,.
\end{align*}
Now inserting our relation of $v$ and $p$ we obtain:
\begin{align*}
v(t^{k+1})
&=\frac{\Delta t^k}{\Delta t^{k+1}} v(t^k) - \frac{\Delta t^{k}}{\Delta t^{k+1}} (u_0-u(t^{k+1})).
\end{align*}
This is the update formula in line 6 of Algorithm~\ref{alg:inv} in the case of variable step sizes. Finally \sout{note that} the discretization in Eq.~\eqref{eq:iss} means that:
$$ t^{k+1} = t^k + \Delta t^k = \sum_{i=0}^k  \Delta t^k \,. $$
Thus, considering \sout{when thinking about/looking for}  quantities on a continuous time scale we \sout{need to} interpret:
$$ u(t^{k+1}) = u(\sum_{i=0}^k  \Delta t^k). $$
Interestingly, the inverse scheme \sout{we used} is extremely forgiving in terms of truncation errors given by the finite number of iterations in Alg.~\ref{PDHG}, and does not accumulate them as proven by Yin and Osher in \shortcite{Yin:2013}. Therefore, one could also consider to trade computational efficiency against error by relying on the latter property of the inverse flow.

\vspace{1ex}\noindent\textbf{The forward flow} (Eq.~\ref{eq:pde}) is given by:
\begin{align*}
    \begin{cases}
    \frac{\partial u}{\partial t}(t) = -p(t), \\
    p(t) \in \partial TV(u(t)), \\ 
    u(0) = u_0.
    \end{cases}
\end{align*}
The implicit Euler discretization of such a flow yields:
\begin{align*}
\frac{u(t^{k+1}) - u(t^k)}{\Delta t^k} = - p(t^{k+1})\Rightarrow \quad & 0 = \frac{1}{\Delta t^k}(u(t^{k+1}) - u(t^k)) + p(t^{k+1})
\end{align*}
which is the optimality condition to:\sout{ the following optimization problem:}
$$ \min_u \frac{1}{2\Delta t^k}\|u - u(t^k)\|_2^2  + TV(u). $$
Comparing this to our Algorithm~\ref{alg:1}, we have to identify $\alpha$ with $\Delta t^k$. Again, the discretization yields:
$$ t^{k+1} = t^k + \Delta t^k = \sum_{i=0}^k  \Delta t^k \,. $$

\subsection{Edge-based gradient}\label{sec:edgegrad}
The gradient operator for piecewise-constant signals defined on mesh triangles can be defined, for each pair of adjacent triangles, as the jump discontinuity across the shared edge. The operator is assembled as a $|E| \times |F|$  matrix $\mathbf{G}_e$, \sout{which has} \textcolor{red}{with} the same zero pattern as the edge-to-triangle adjacency matrix; it contains $\pm 1$ as its values, s.t. each row sums up to zero.
The divergence operator is defined as:
\sout{The relative divergence operator is defined as the negative adjoint operator:}
\begin{align*}
    \mathbf{D}_e = -\mathbf{T}^{-1}\mathbf{G}_e^\top \mathbf{A}_e\,,
\end{align*}
where $\mathbf{A}_e$ is a diagonal matrix of edge lengths, \sout{playing the role of}i.e. the area elements for this function space, and  $\mathbf{T}$ is the matrix of triangle areas.

\sout{This gradient corresponds to the discrete differential operator of $1$-forms in the language of discrete exterior calculus desbrub:2005, while the divergence operator definition follows from applying Stokes' theorem, where $\mathbf{A}_e $  and $\mathbf{T}$ play the role of the Hodge star operator.} 
%

\subsection{Parameters and stability}\label{sec:params}
We expose four parameters: the step size $\alpha$, the number of spectral components $N$ (both used in Algorithm~\ref{alg:1}), and the interior step sizes $\sigma,\tau$ in Algorithm~\ref{PDHG}. \sout{In practice, the only important parameter is $\alpha$, since $N$ can be set automatically as a function of $\alpha$.}\textcolor{red}{The only parameter which requires tuning is $\alpha$, while the others can be set automatically, accordingly.}



\vspace{1ex}\noindent\textbf{Step size} $\alpha$. Since Algorithms~\ref{alg:1} and \ref{alg:inv} are {\em unconditionally stable}, we can choose arbitrarily small or large values for $\alpha$. This determines the resolution of the spectral representation; to get a good separation of features, one should increment $\alpha$ according to a non-uniform sampling of the spectral domain, with higher sampling density within the bands having stronger spectral activity. 
In practice, following Algorithm~\ref{alg:inv}, we proceed by estimating the maximum diffusion time by choosing the smallest value for $\alpha$ that makes the fidelity term negligible in the following energy:
\begin{align*}
\frac{1}{2\alpha}\int_\mathcal{M}(u(x)-\mathrm{mean}(u_0))^2 dx+ TV(u)\,.
\end{align*}
For such an $\alpha$, minimizing the energy above corresponds to minimizing the total variation alone. \sout{In practice,} This ensures that no spectral activity exists beyond the diffusion interval $(0,\alpha)$.
Once the maximum possible $\alpha$ is estimated, we rescale this value as $\alpha\xleftarrow\ C \alpha$, at each iteration (with $C \in[0,1]$, typically $C=0.7$). \sout{(we decrease $\alpha$ instead of increasing it, since in our tests we use the inverse scale space flow in Algorithm~\ref{alg:inv} instead of Algorithm~\ref{alg:1}). This kind of rescaling implements a non-uniform sampling of the spectrum and ensures that more samples are placed in correspondence with high-frequency content.}
Decreasing $\alpha$ by a different schedule \sout{than $\alpha\xleftarrow\ 0.7\alpha$} would still ensure a perfect reconstruction of the input signal; no details are ever lost in the analysis and synthesis process. Instead, the specific choice affects the {\em resolution} at which the spectral components are extracted, and depends on the specific application, as briefly discussed in the limitations paragraph of Section~\ref{sec:limitations}.

\vspace{1ex}\noindent\textbf{Number of spectral components} $N$. This parameter can be determined automatically, since it is equivalent to the number of rescalings of $\alpha$ required to cover the interval $(0,\alpha)$, assuming $\alpha$ is chosen according to the strategy above.


\vspace{1ex}\noindent\textbf{Interior step sizes} $\sigma,\tau$. \sout{These influence the runtime of the algorithm.} To ensure stability, these parameters must be chosen so as to satisfy:
\begin{align*}
     \tau \sigma \leq \frac{\min(\|e\| \in E)}{\|\mathbf{G}\|^2_{o}}\,,
\end{align*}
where $\| \mathbf{G} \|_{o}= \sup \left\{\frac{\|G u\|}{\|u\|} \ u \in \mathcal{L}^2(\M)\right\}$ is the operator norm, and the numerator denotes the minimum edge length on the mesh. We estimate $\| \mathbf{G} \|_{o}$ with the \texttt{normest} function in Matlab and set $\sigma=\tau$.
´
   
\vspace{1ex}\noindent\textbf{Mesh quality.}
Like most geometry processing algorithms, our spectral decomposition is susceptible to \sout{crude and} low-quality input meshes. 
 \sout{that are often found ``in the wild.''} \sout{Such} Tiny angles cause an increase in the gradient norm $\|\mathbf{G}\|_{o}$, leading \sout{in turn} to a decrease in convergence speed of the diffusion process.
\begin{wrapfigure}[11]{r}{0.5\linewidth}
\begin{center}
\begin{overpic}
[trim=0cm -3.9cm 0cm 0cm,clip,width=0.3\linewidth]{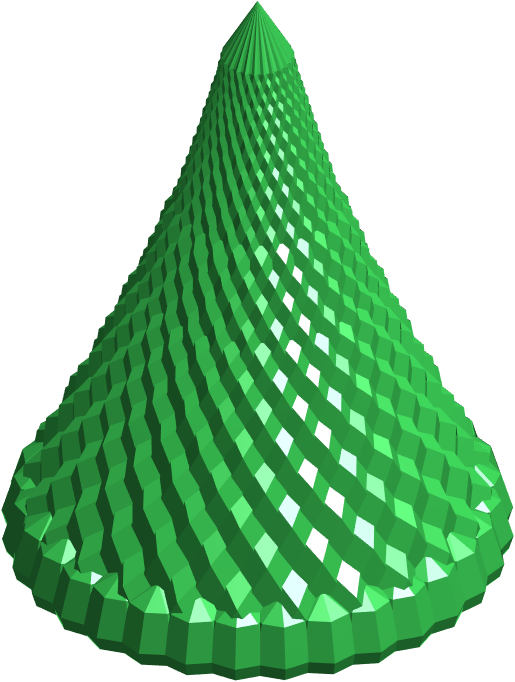}
\end{overpic}\hspace{0.04cm}
\includegraphics[width=0.6\linewidth]{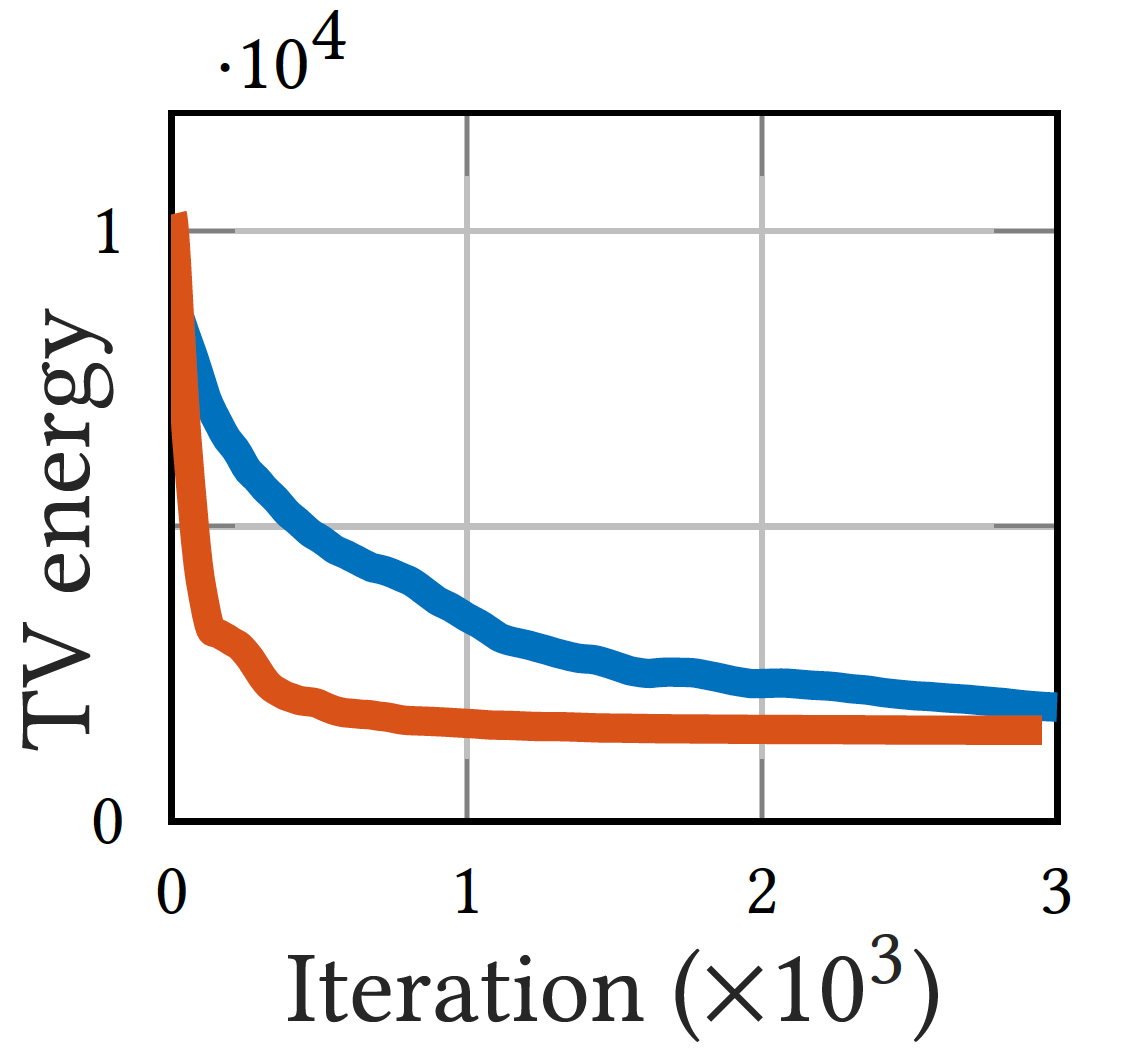}
\begin{overpic}
[trim=0cm 0cm 0cm 0cm,clip,width=0.9\linewidth]{./images/treeside.png}
\end{overpic}
\end{center}
\end{wrapfigure}
To address this, we adopt the \sout{intrinsic triangulation} approach of \cite{sharp2019navigating}, which decouples the triangulation used to describe the domain from the one used in the decomposition algorithm. \textcolor{red}{In the inset figure, we show an example where degenerate triangles (visible from the bottom of the drill, bottom row) cause distortion when we run the TV flow on the normal field (blue curve). With intrinsic triangulations (red curve), we get the expected results.}

\sout{In practice, this boils down to constructing our operators on top of an alternative data structure. In the inset we show an example in which we run the TV flow on the normal field $\vec{n}:\M\to\mathbb{R}^3$ of the drill shape (top left) as input signal. This shape is very problematic, since it has a large percentage of degenerate triangles and a high density of tiny angles at the basis (see the bottom-left figure for a view from the side and from the bottom; the tiny angles show as a black blot). We plot energy vs time for two cases: with (red) and without (blue) intrinsic triangulations. The resulting shapes at convergence are shown in the bottom-center and bottom-right (without and with intrinsic triangulations respectively). We see that intrinsic triangulations help reach convergence more quickly and also avoid numerical instabilities that occur when a low-quality mesh is used.}

\section{Runtime statistics}\label{sec:runtime}
The time performance of the spectral decomposition depends on a few factors: complexity of the domain, \sout{number of internal} time steps, and number of spectral components to compute.
\sout{We start by considering} A single iteration of Alg.\eqref{PDHG} \sout{the PDHG algorithm (line 5 in Algorithm~\ref{alg:inv})} \sout{. The algorithm} has a convergence rate of $\frac{1}{k^2}$ w.r.t. the primal-dual gap (see \cite{chambolle:2010}). The number of iterations \sout{needed for taking} \textcolor{red}{to bring} the gap error below a desired threshold however \sout{largely} depends on the internal step sizes \sout{parameters} $\sigma, \tau$ \sout{, which, as stated in} (see Appendix~\ref{sec:params})\sout{, must obey a precise condition in order to ensure stability}. 
%
To provide a good intuition for the actual performance \sout{of our method} we report in Table~\ref{tab:times} the \sout{computational} runtimes \sout{needed} for some of the examples showcased in the paper. The computations were done in Matlab with an unoptimized single-threaded implementation, setting the error threshold of PDHG to $10^{-4}$ and a maximum number of iterations to $1000$ and $300$, for scalar and vector signals respectively.
\begin{table}
\resizebox{0.75\linewidth}{!}{
  \begin{tabular}{lllllc}
    \toprule
    \textbf{Example} & \textbf{|V|}& \textbf{|F|}&\textbf{N} & \textbf{Time} & \textbf{Rec Time} \\
    \midrule
    Figure \ref{fig:duck}& 6079  & 2000 & 50   & 12.9 s & -\\
    Figure \ref{fig:shirt}& 2590  & 2000 & 40  & 3.9 s & -\\
    Figure \ref{fig:removal}& 23308  & 45953 & 30   & 16.59 s &  0.16 s\\
    Figure \ref{fig:totem}& 44313  & 88622 & 25   & 31.27 s &  0.46 s\\
    Figure \ref{fig:teaser}& 157056 & 314112  & 30   & 180.5 s & 2.50 s \\
    Figure \ref{fig:lion}& 50002 & 100000 & 20   & 23.69 s & 0.41 s\\
    Figure \ref{fig:gargoyle}& 50002  & 100000 & 20   & 28.5 s & 0.52 s\\
  \bottomrule
\end{tabular}}
\caption{\label{tab:times}Table of computational times for spectral TV. Left to right: input, number of vertices, number of faces, number of spectral components computed, total time for computing the spectral decomposition, total time for recovering vertex coordinates, when needed (i.e., solution of Eq. \eqref{eq:poisson}).}
\end{table}

\end{document}